\documentclass{article}
\usepackage[english]{babel}
\usepackage{amsmath}
\usepackage{amsfonts}
\usepackage{graphicx}
\usepackage[colorlinks=true, allcolors=blue]{hyperref}
\usepackage{comment}
\usepackage{enumitem}
\usepackage{mathtools}
\usepackage{amssymb}
\usepackage{amsthm}
\usepackage{subcaption}
\usepackage{textcomp}
\usepackage[table]{xcolor}

\usepackage {tikz}
\usetikzlibrary {positioning}
\usetikzlibrary{patterns}
\tikzset{main node/.style={circle,fill=blue!20,draw,minimum size=2cm,inner sep=0pt},}

\usepackage{array}
\newcolumntype{P}[1]{>{\centering\arraybackslash}p{#1}}

\newcommand{\vect}[1]{\boldsymbol{#1}}

\title{Thermodynamic fluctuation theorems govern human sensorimotor learning}
\author{P. Hack, C. Lindig-Leon, S. Gottwald, D. A. Braun}
\date{ }

\begin{document}
\maketitle
\begin{abstract}
The application of thermodynamic reasoning in the study of learning systems has a long tradition. Recently, new tools relating perfect thermodynamic adaptation to the adaptation \emph{process} have been developed. These results, known as fluctuation theorems, have been tested experimentally in several physical scenarios and, moreover, they have been shown to be valid under broad mathematical conditions. Hence, although not experimentally challenged yet, they are presumed to apply to learning systems as well. Here we address this challenge by testing the applicability of fluctuation theorems in learning systems, more specifically, in human sensorimotor learning. In particular, we relate adaptive movement trajectories in a changing visuomotor rotation task to fully adapted steady-state behavior of individual participants. We find that human adaptive behavior in our task is generally consistent with fluctuation theorem predictions and discuss the merits and limitations of the approach.  \end{abstract}


\section{Introduction}
The study of learning systems with concepts borrowed from statistical mechanics and thermodynamics has a long history reaching back to Maxwell's demon and the ensuing debate on the relation between physics and information \cite{parrondo2015thermodynamics}. Over the last 20 years, the informational view of thermodynamics has experienced great developments, which has allowed to broaden its scope form equilibrium to non-equilibrium phenomena \cite{jarzynski2011equalities,de2013non}. Of particular importance are the so-called fluctuation theorems \cite{seifert2012stochastic,jarzynski2000hamiltonian,crooks1999entropy}, which relate equilibrium quantities to non-equilibrium trajectories allowing, thus, to approximate equilibrium quantities via experimental realizations of non-equilibrium processes \cite{ytreberg2004efficient,park2003free}. Among the fluctuation theorems, two results stand out, Jarzynski's equality \cite{jarzynski1997equilibrium,cohen2004note,jarzynski2004nonequilibrium} and Crooks' fluctuation theorem \cite{crooks1998nonequilibrium,crooks2000path}, as they aim to bridge the apparent chasm between reversible microscopic laws 
and irreversible macroscopic phenomena \cite{loschmidt1876ueber}.

The advances in non-equilibrium thermodynamics have recently also led to new theoretical insights into simple learning systems \cite{goldt2017stochastic,perunov2016statistical,england2015dissipative,still2012thermodynamics,ortega2013thermodynamics,grau2018non}.
Abstractly, thermodynamic quantities like energy, entropy or free energy can be thought to define order relations between states \cite{lieb1991,gottwald2019}, which makes them applicable to a wide range of problems. In the economic sciences, for example, such order relations are typically used to define a decision-maker's preferences over states \cite{mascollel1995}. Accordingly, a decision-maker or a learning system can be thought to maximize a utility function, analogous to a physical system that aims to minimize an energy function.
Moreover, in the presence of uncertainty in stochastic choice, such decision-makers can be thought to operate under entropy constraints reflecting the decision-maker's precision \cite{ortega2013thermodynamics,parrondo2015thermodynamics}, resulting in soft-maximizing the corresponding utility function instead of perfectly maximizing it. This is formally equivalent to following a Boltzmann distribution with energy given by the utility. Therefore, in this picture, the physical concept of work corresponds to utility changes caused by the environment, whereas the physical concept of heat corresponds to utility gains due to internal adaptation \cite{still2012thermodynamics}. Like a thermodynamic system is driven by work, such learning systems are driven by changes in the utility landscape (e.g. changes in an error signal). By exposing learning systems to varying environmental conditions, it has been hypothesized that adaptive behavior can be studied in terms of fluctuation theorems \cite{grau2018non,england2015dissipative}, which are not necessarily tied to physical processes but are broadly applicable to stochastic processes satisfying certain constraints \cite{hack2022jarzyskis}.

Fluctuation theorems are usually deployed in statistical mechanics; particularly, the study of nonequilibrium steady states in thermodynamics. In this setting, one normally assumes a probabilistic description of an ensemble of many particles, i.e., the kinds of systems usually considered in statistical thermodynamics. However, as described in \cite{seifert2005entropy,seifert2012stochastic}, exactly the same principles and fluctuation theorems also apply to the path of a single particle, leading to stochastic thermodynamics. This suggests that fluctuation theorems may not only be applicable to the statistics of ensembles of many learners, but also when describing the trajectory of a single participant during a learning process.

Although fluctuation theorems have been empirically observed in numerous experiments in the physical sciences \cite{douarche2005experimental,collin2005verification,saira2012test,liphardt2002equilibrium,an2015experimental,smith2018verification},  there have been no reported experimental results relating fluctuation theorems to adaptive behavior in humans or other living beings. Here, we test Jarzynski's equality and Crooks'  fluctuation theorem experimentally in a human sensorimotor adaptation task.
In this context, the fluctuation theorem establishes a linear relationship between the externally imposed utility changes driving the learning process (which are directly related to non-predicted information and energy dissipation \cite{still2012thermodynamics}) and the log-probability ratio between forward and backward adaptation trajectories, when exposing participants to the sequence of environments either in the forward or reverse order. Accordingly, such learners can be quantitatively characterized by a hysteresis effect that can also be observed in simple physical systems.

\section{Results}
\label{sect:results}

\begin{figure}[!tb]
\centering
    \includegraphics[scale=.32]{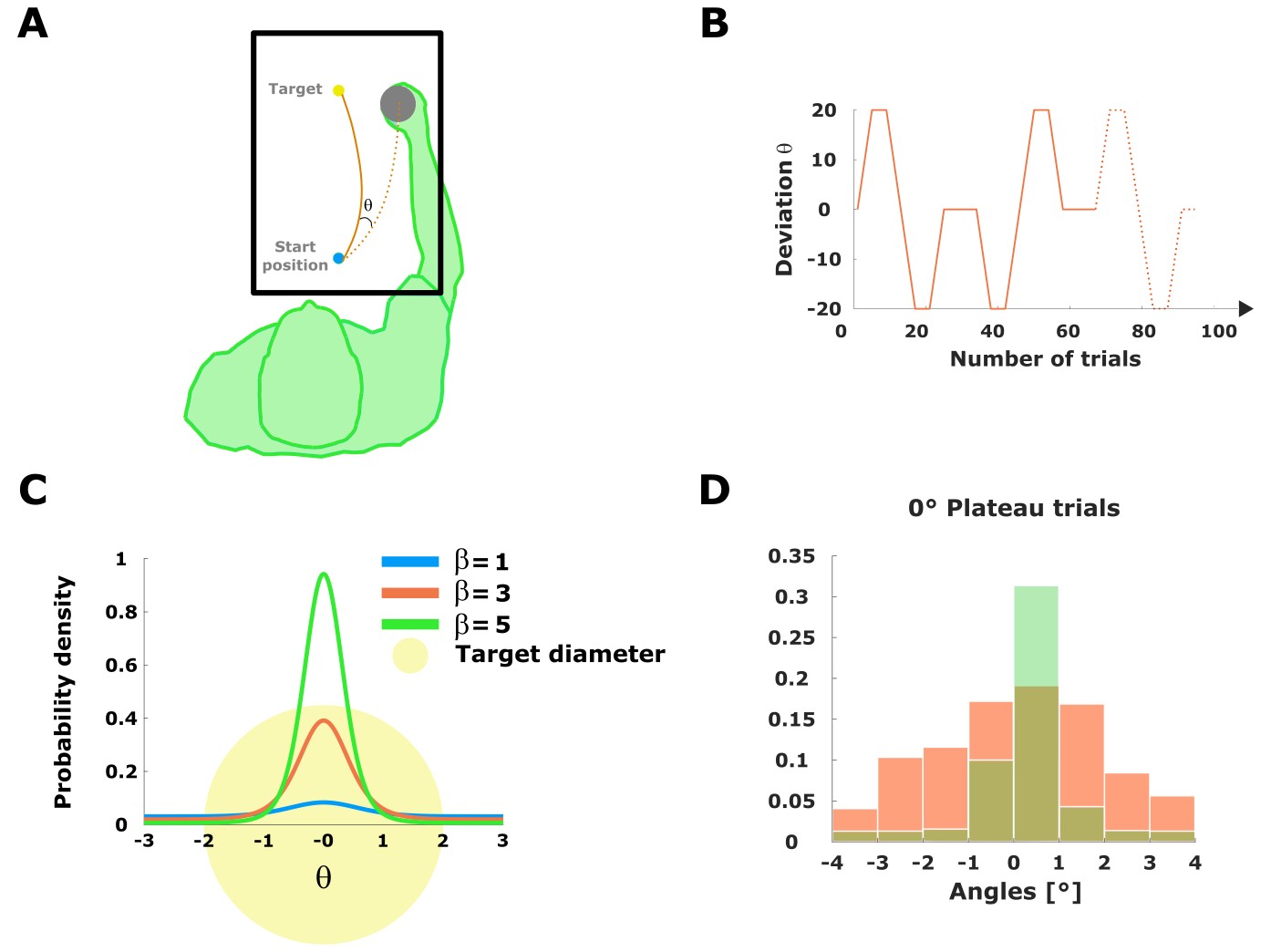}
    \caption{\textbf{A} Schematic representation of an experimental trial with deviation angle $\theta$. The dotted line represents the participant's hand movement and the continuous line represents the rotated movement observed on the screen. \textbf{B} Experimental protocol. The continuous line represents the deviation angles $\theta$ imposed during one experimental cycle, where trials 1 to 25 constitute the forward process and trials 34 to 58 constitute the backward process. The dotted line represents the beginning of the next cycle. \textbf{C} Illustration of the equilibrium distributions \eqref{Boltzmann} with $b,\theta_n=0$ resulting from the exponential quadratic error \eqref{utility} and, respectively, $\beta=1,1.5,2$.
    The shaded area represents the target, which tolerates, at most, an error of $2 ^\circ$. \textbf{D} Comparison between the equilibrium distributions that we fit using the initial 100 trials (before participants experience any perturbation) and participants’ performance in the washout plateaus between cycles (the sequence of trials with $\theta = 0$ that separate forward and backward protocol), to check whether participants equilibrate between cycles, as required by the fluctuation theorem.
    Red shows the normalized error histogram for the in-between plateaus exemplarily for participant 7, green shows the histogram of the fitted equilibrium distribution for the initial block of 100 trials of the same participant.} The comparison for all other participants can be found in Figure \ref{plateaus all}.
    \label{methods exp}
\end{figure}

In a visuomotor adaptation task, human participants controlled a cursor on a screen towards a single stationary target by moving a mechanical manipulandum that was obscured from their vision under an overlaid screen---see Figure~\ref{methods exp}\textbf{A}. Crucially, in each trial $n$, the  position  of  the  cursor  could be rotated with angle $\theta_n$ relative to the actual hand position so that participants had to adapt when moving the cursor from the start position to the target. To measure participants' adaptive state, we recorded their movement position at the time of crossing a certain distance from the start position, so that  their response could be characterized by an angle $x_n$. 
The deviation between participants' response $x_n$ and the required movement incurs a sensorimotor loss $E_n$ \cite{kording2004loss} in trial $n$, that can be quantified as an exponential quadratic error 
\begin{equation}
	\label{utility}
	E_n(x)= 1 - e^{- (x-(\theta_n+b))^2},
\end{equation}
that depends on the actual rotation angle  $\theta_n$ set in trial $n$. The parameter $b$ is a participant-specific parameter allowing for bias due to posture, biomechanics, the mechanics of the manipulandum, or other influences---see Figure \ref{methods exp}\textbf{D}. The loss~\eqref{utility}  is taken to be the energy (or negative utility) of a participant's stochastic response $X_n = x_n$. For a bounded rational decision-maker \cite{ortega2013thermodynamics,Schach2018,cecilia2019,cecilia2021} that optimizes this loss under uncertainty,  
the optimal pointing behavior after a suitably long adaption time is described by a Boltzmann equilibrium distribution $p_n^{eq}$ of the form
\begin{equation}
\label{Boltzmann}
    p_{n}^{eq}(x_n) =  \exp\big(-\beta( E_n(x_n) - F_n )\big),
\end{equation}
for all $x_n\in A_n$, where the sensorimotor error $E_n(x_n)$ plays the role of an energy, the free energy term $F_n = \frac{1}{\beta} \log \int_{A_n} \exp\left( -\beta E_n(x_n) \right) dx_n$ is caused by the normalization, and $A_n$ is the support of the equilibrium distribution $p_{n}^{eq}$, which will vary for each participant, as we explain in Section \ref{exp design}. See Figure \ref{methods exp} \textbf{C} for a representation of \eqref{Boltzmann}. Moreover, the softness-parameter $\beta$, also known as \textit{inverse temperature} or \emph{precision}, controls the trade-off between entropy maximization and energy minimization, essentially interpolating between a purely stochastic choice ($\beta = 0$) and a purely rational choice ($\beta \to \infty$) minimizing the energy perfectly. 

The task consisted of a sequence of target reaching trials, where the rotation angle $\theta_n$ changed from one trial $n$ to the next trial $n+1$ according to a given up-down protocol---see Figure~\ref{methods exp}\textbf{B}---, so that participants' responses over trials could be represented by a trajectory  $\vect{x}=(x_0,x_1,..,x_N)$.
When the environment is changing over trials, we can distinguish cumulative error changes \smash{$\Delta E_{ext}(\vect{x}) \coloneqq \sum_{n=0}^{N-1} (E_{n+1}(x_n)-E_{n}(x_n))$} that are induced externally by changes in the environmental parameter $\theta_n$, from cumulative error changes \smash{$\Delta E_{int}(\vect{x}) \coloneqq \sum_{n=1}^{N} (E_{n}(x_n)-E_{n}(x_{n-1}))$} due to internal adaptation when subjects change their response from $x_{n-1}$ to $x_{n}$. Crucially, it is exactly the externally induced changes in error, $\Delta E_{ext}(\vect{x})$, analogous to the physical concept of work, that drive the adaptation process: if $\Delta E_{ext}(\vect{x})$ is large, the system is more surprised and has to adapt more. In the following, we thus refer to $\Delta E_{ext}(\vect{x})$ as \emph{driving error} or \emph{driving signal}. When applying Crooks' fluctuation theorem for general adaptive systems \cite{hack2022jarzyskis} to the above setting, we obtain the linear relation
\begin{equation}
	\label{crooks}
	\Delta E_{ext}(\vect{x}) - \Delta F = \frac{1}{\beta}\log \left( \frac{\rho^F(\vect{x})}{\rho^B(\vect{x}^R)} \right),
\end{equation}
where $\vect{x}^R = (x_N,\ldots,x_1)$ is the reverse trajectory, $\Delta F$ denotes the free energy difference $F_N-F_0$ and the distributions $\rho^F(\cdot)$ and $\rho^B(\cdot)$ denote the probability of observing a certain trajectory when the learner faces a series of environments in some specific order or the order is reversed, respectively. This form of Crooks' theorem allows for an intuitive interpretation, in that any difference in probability of a trajectory and its reverse signifying a hysteresis can be directly related to an excess loss that is irretrievably generated because of imperfect adaptation. Unfortunately, Equation~\eqref{crooks} is hard to determine from data, as it would require to estimate probability distributions over paths. However, there is an equivalent form of Crooks' theorem that groups all trajectories according to their associated value of $\Delta E_{ext}(\vect{x})$ with corresponding distributions   $\rho^F$ and $\rho^B$ over these values, such that
\begin{equation}
	\label{prediction}
	\Delta E_{ext}(\vect{x}) - \Delta F = \frac{1}{\beta}\log \left( \frac{\rho^F(\Delta E_{ext}(\vect{x}))}{\rho^B(-\Delta E_{ext}(\vect{x}))} \right).
\end{equation}
The distribution $\rho^F(\cdot)$ can be interpreted as the probability that the learner experiences a certain overall surprise when being exposed sequentially to a series of environments and $\rho^B(\cdot)$ is the analogous concept when the order in which the environments are presented is reversed. In equation \eqref{prediction}, these densities are evaluated at the actual driving errors $\Delta E_{ext}(\vect{x})$ and $-\Delta E_{ext}(\vect{x})$, respectively, for a particular adaptive trajectory $\vect{x}$.

A direct consequence of \eqref{prediction} is  Jarzynski's equality \cite{crooks1998nonequilibrium}, which states that
\begin{equation}
\label{prediction II}
    \big\langle e^{-\beta \Delta E_{ext}(\vect{X})}\big\rangle = e^{-\beta \Delta F},
\end{equation}
where $\langle ~ \cdot ~ \rangle \coloneqq \mathbb E[~\cdot~]$ denotes the expectation operator, considering $\vect{X} = (X_n)_{n=0}^N$ a Markov chain with transition densities $\Pi_n$ that have $p^{eq}_n$ as stationary distributions, that is, for each $n$, $p^{eq}_n$ is the stationary distribution for $X_n$. In our experiment, $\vect{X}$ represents participants' responses that are repeated over multiple repetitions of the forward-backward protocol.    
In the following, we will test the relationships \eqref{prediction} and \eqref{prediction II} experimentally with $\Delta F = 0$ as our human learners start and end in the same environmental state (i.e. $F_N=F_0$). Note that, in our particular setting where there is no overall change in the free energy $(\Delta F=0)$, Equation~\eqref{prediction II} suggests that the expected value $\big\langle e^{-\beta \Delta E_{ext}(\vect{X})}\big\rangle$ equals $e^{-\beta 0}=1$ irrespective of the value taken by $\beta$. This provides a quantitative prediction that we will evaluate empirically below.


In our experiment the task is divided into 20 cycles of 66 trials each, following the protocol  \eqref{protocol}  illustrated in  Figure \ref{methods exp}\textbf{B}. We refer to trials 1 to 25 of each cycle as a realization of the \emph{forward process} and trials 34 to 58 as a realization of the \emph{backward process}. Notice the backward process consists of the same angles as the forward process, that is, the same utility functions, but in reversed order. Thus, we record for each participant 20 values for $\Delta E_{ext}(\vect{x})$ in both the forward and backward processes
that we use to estimate participants' probability densities of the forward and backward processes, $\rho^F$ and $\rho^B$, respectively, using kernel density estimation. As the amount of data is limited to test the linear relation in \eqref{prediction}, we will use simulation results in the following to compare against participants' behavior.

\begin{figure}[!tb]
\centering
    \includegraphics[scale=.2]{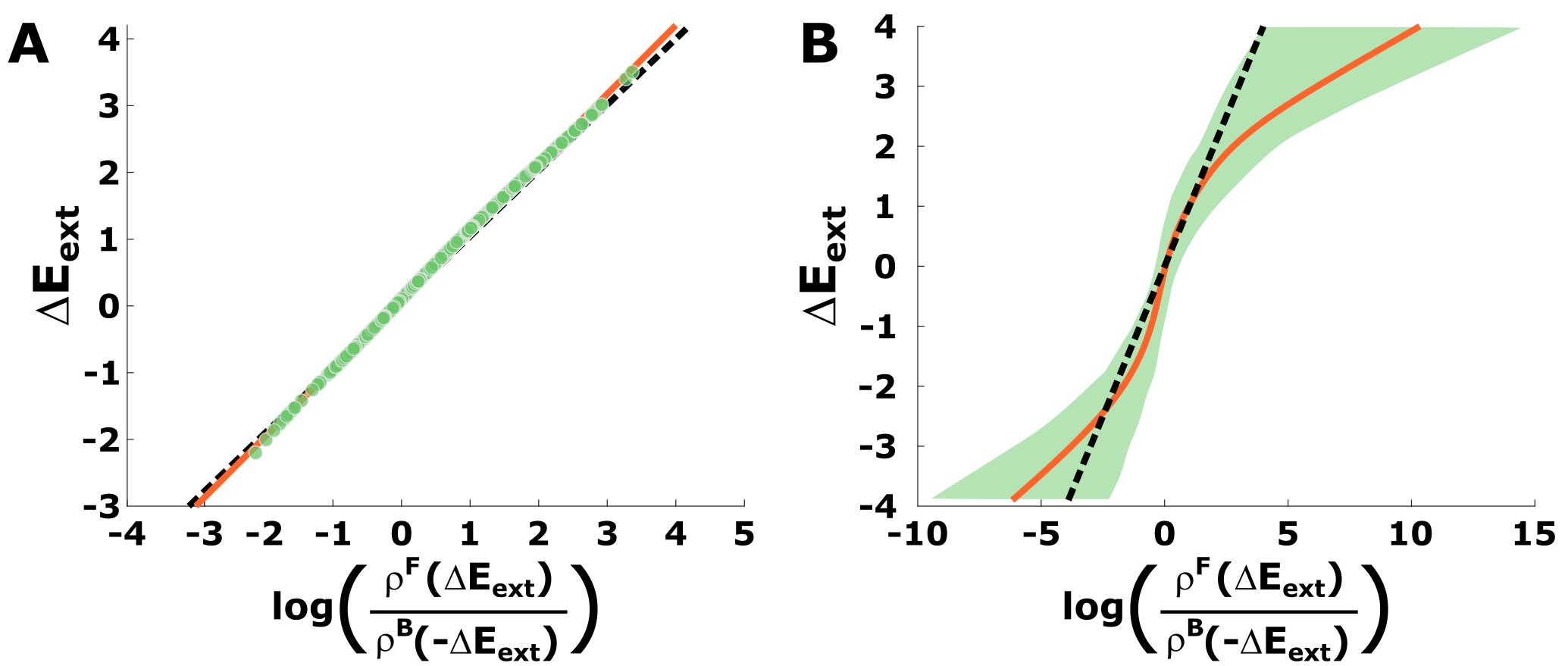}
    \caption{Simulation of Crooks' fluctuation theorem. \textbf{A} Simulation with 1000 cycles. In black, the theoretical prediction; in red, the linear regression for the simulated data and, in green, the simulated points. Since the simulated data set adjusts pretty well to Crooks' fluctuation theorem \eqref{prediction}, Jarzynski's equality \eqref{prediction II} is fulfilled. \textbf{B} Simulation with 20 cycles and bootstrapping. The black line is the theoretical prediction \eqref{prediction} while the red line and shaded area are, respectively, the mean and the 99 \% confidence interval of \eqref{prediction} after 1000 bootstraps of the driving error values obtained in a single run (which consists of 20 cycles).}
    \label{2 simus}
\end{figure}

When simulating an artificial decision-maker based on a stochastic optimization scheme with Markovian dynamics, for example a Metropolis-Hasting algorithm with target distribution $p_n^{eq}\propto \exp(-\beta E_n)$, it is clear that we can recover the linear relationship \eqref{prediction}, provided that sufficient samples are collected \cite{hack2022jarzyskis}---see, for example, a simulation with 1000 cycles in Figure \ref{2 simus}\textbf{A} where we can see a good adjustment between the theoretical prediction (in black) and the linear regression of the observed data (in red). As a result, \eqref{prediction II} also holds in this scenario. The more critical question is what happens when only few samples are available. To this end, we use the stochastic optimization algorithm to simulate the protocol of our experiment, that is, 20 cycles, and indicate confidence intervals using 1000 bootstraps. It can be seen in Figure \ref{2 simus}\textbf{B} that the theoretical prediction is consistent with the $99\%$ confidence interval in the region where $|\Delta E_{\textrm{ext}}| \leq 4$ (which is the region where our experimental data lies).
Using the same bootstrapped data, we obtain several estimates of $\langle e^{- \Delta E_{ext}(\vect{X})}\rangle$ (the mean of $e^{- \Delta E_{ext}(\vect{X})}$ for the observed values of $\Delta E_{ext}(\vect{X})$ at each bootstrap) which we use to calculate a confidence interval for it. This results in the $99\%$ confidence interval for $\langle e^{-\Delta E_{ext}(\vect{X})}\rangle$ being $(0.48,\text{ }1.64)$, which is consistent with the theoretical prediction $\langle e^{-\Delta E_{ext}(\vect{X})}\rangle = 1$ for $\Delta F = 0$ according to Equation~\eqref{prediction II}. Accordingly, we will expect a similar behavior for our experimental data. Note we take, for simplicity, $b=0$, $\beta=1$ and, for all $n$, $A_n=[-90,90]$ in these simulations (see Methods).

\begin{figure}[!tb]
\centering
    \includegraphics[scale=0.21]{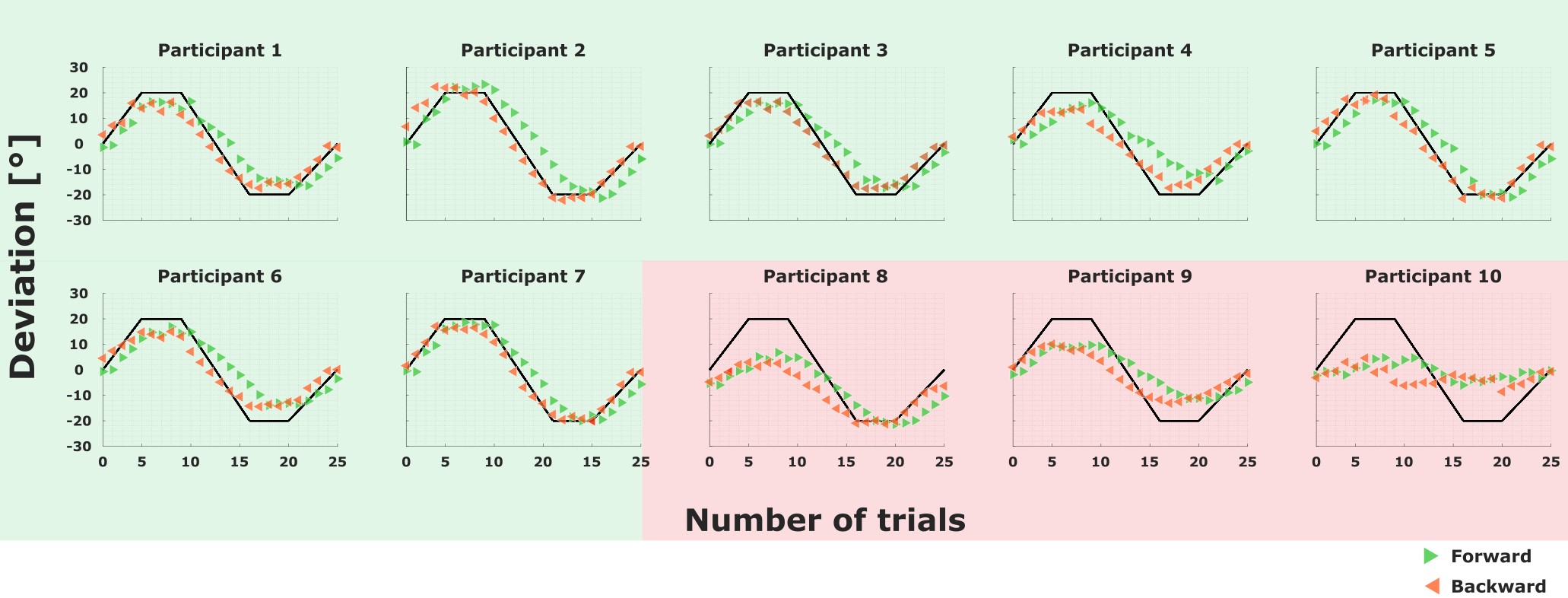}
    \caption{Hysteresis effect. The filled triangles are the mean of the observed angles for every deviation in both the forward process, in green, and the backward process, in red. The black line is the forward protocol. Note that we have mirrored the triangles for the backward process to make them coincide with those in the forward process that are exposed to the same true angle. Participants that achieve at least $50\%$ adaptation are shaded by a green background color. Hysteresis can be observed between trials 1 and 5, 9 and 17 and 21 and 25. Notice,  as expected, the forward means are below the backward in the first region, above in the second and below again in the third.}
    \label{hyste plot}
\end{figure}

Participants' average adaptive responses can be seen in Figure \ref{hyste plot} compared to the experimentally imposed true parameter values (the trial-by-trial responses can be seen in Figure \ref{forward all}). The green and red lines distinguish the forward and backward trajectories, respectively, so that, from the contrast between the two curves, hysteresis becomes apparent, as common in simple physical systems \cite{jarzynski2011equalities} and as reported previously in similar experiments for sensorimotor adaptation \cite{turnham2012facilitation}. Participants that achieve at least $50\%$ adaptation are shaded by a green background color and are our participants of interest. The three participants that fail to achieve this minimum adaptation level are marked by a red shade. Instead of excluding these participants entirely from the analysis, we keep them in to show the contrast to the well-adapted participants and to highlight that the results reported for the well-adapted participants do not hold trivially for any participant producing inconsistent behavior.

\begin{figure}[!tb]
\centering
    \includegraphics[scale=0.21]{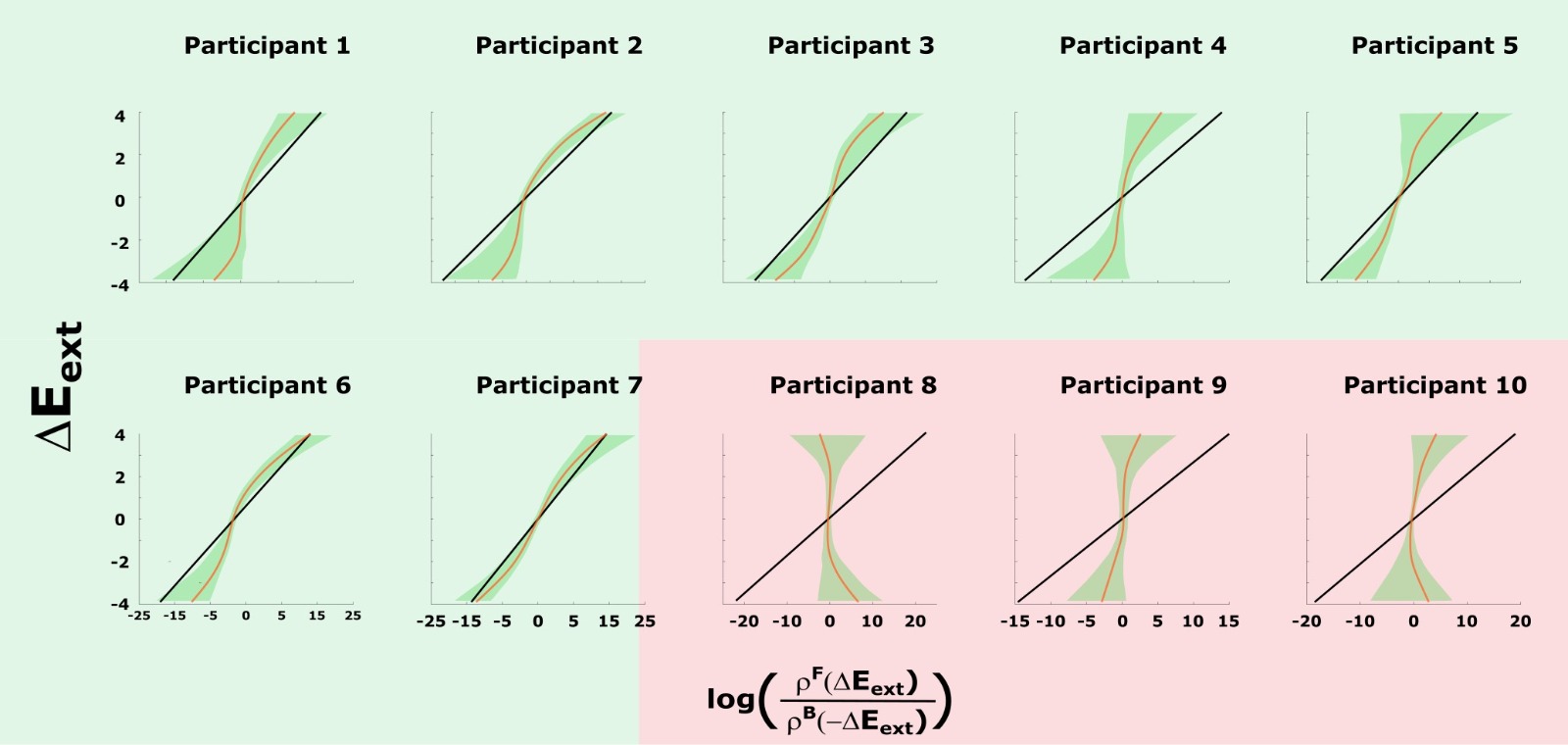}
    \caption{Experimental results for Crooks' fluctuation theorem when the sensorimotor loss behaves as an exponential quadratic error \eqref{utility}. The black line is the theoretical prediction of Crooks' fluctuation theorem \eqref{prediction} while the curves stand for the mean path after 1000 bootstraps of the observed driving error values. Participants that achieve at least $50\%$ adaptation (c.f. Figure \ref{hyste plot}) are shaded by a green background color. The shaded areas inside the graphs are the 99\% confidence intervals which result from bootstrapping. Note we fit the parameters for each participant according to Section \ref{exp design}.}
    \label{together}
\end{figure}

\begin{figure}[!tb]
\centering
    \includegraphics[scale=0.21]{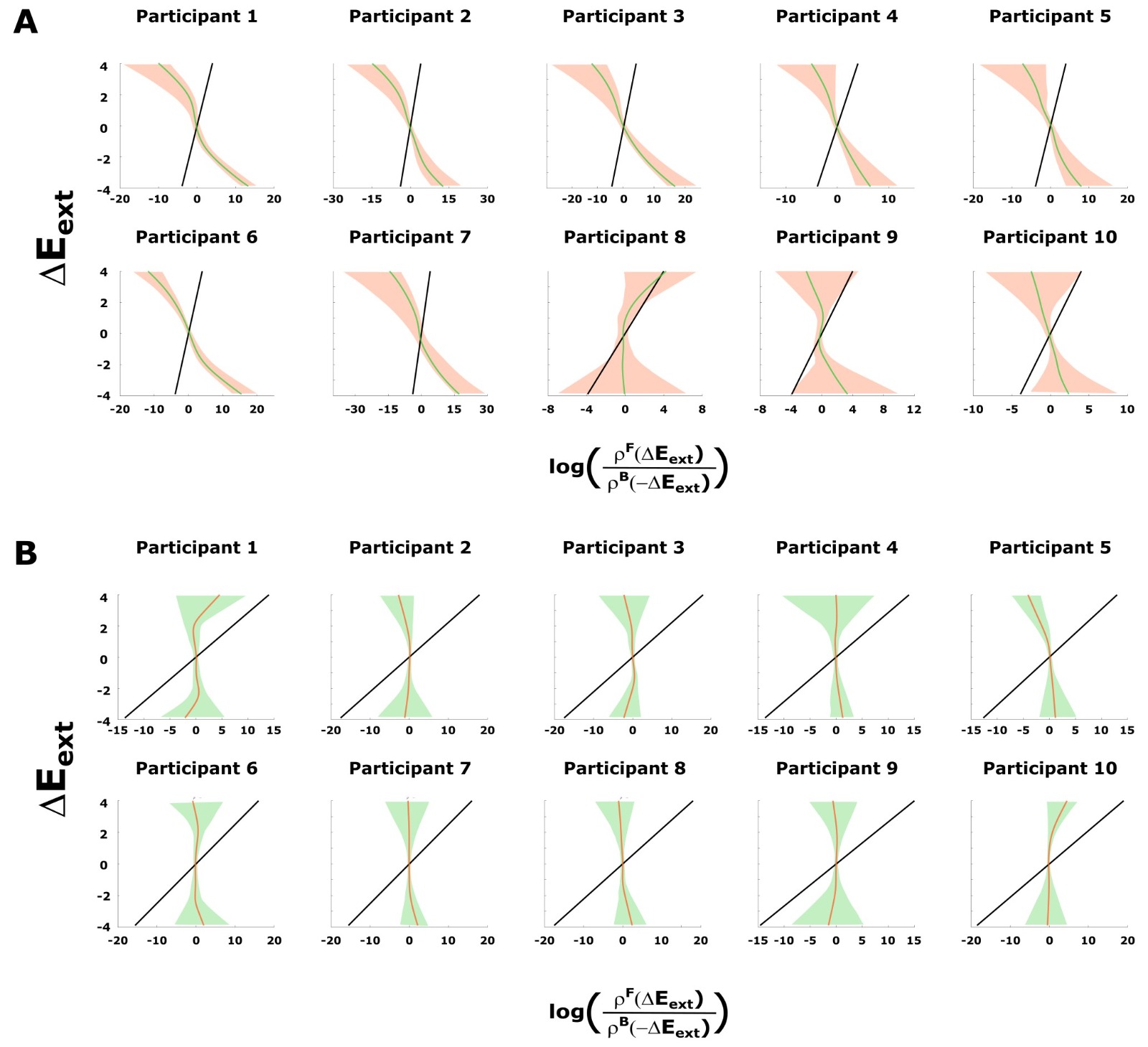}
    \caption{Control results for Crooks' fluctuation theorem in two scenarios: \textbf{A} the sensorimotor loss behaves like a Mexican hat function and \textbf{B} the sensorimotor loss behaves as an exponential quadratic error but we sample the observed angles randomly with repetition. The black line is the theoretical prediction of Crooks' fluctuation theorem \eqref{prediction} while the curves stand for the mean path after 1000 bootstraps of the observed driving error values. The shaded areas inside the graphs are the 99\% confidence intervals which result from bootstrapping. Note, for simplicity, we assume $\beta=1$ for all participants when using the Mexican hat to demonstrate that the result in (A) does not trivially hold for any cost function. For \textbf{B}, we fit the parameters for each participant according to Section \ref{exp design}.}
    \label{togetherB}
\end{figure}

Figure \ref{together} shows participants' data compared to the theoretical prediction from \eqref{prediction} and the 99 \% confidence interval after 1000 bootstraps as in the case of the simulations in Figure \ref{2 simus}\textbf{B}.
There, we see that our data follow the trend of the theoretical prediction and lie within or close to the confidence interval bounds of the prediction in broad regions for several participants.
This is not a trivial result, as can be easily seen, when randomizing the temporal order of the trajectory points or when replacing the utility function with another one that does not fit the setup. Figure \ref{togetherB}\textbf{A} and \ref{togetherB}\textbf{B} show this, for example, for an inverted Mexican hat (\eqref{mex hat} with $\sigma=4$) that assigns low utility to the target region, and for resamples of the trajectory points in a random order, respectively. Both results are clearly incompatible with the theoretical prediction.

When conducting an additional robustness analysis in Figure~\ref{graph distances}, we found that, under the proposed utility function, participants' behavior is compatible with Crooks' fluctuation theorem for a broad neighbourhood of parameter settings, but breaks down when choosing implausible parameters. Regarding Jarzynski's equality \eqref{prediction II}, the confidence intervals for the majority of participants are consistent with the theoretical prediction when using the bootstrapped values to calculate $\langle e^{-\beta \Delta E_{ext}(\vect{X})}\rangle$ (cf. Table \ref{jarz participants}). In contrast, when following the same procedure for both the inverted Mexican hat and the randomized procedure, we obtain consistency for a considerably smaller number of participants. In particular, for the inverted Mexican hat, we obtain consistency for only two participants. Moreover, these participants are $S_8$ and $S_9$, which belong to the group that did not reach at least $50\%$ adaptation (indicated by the red background area in the figures). For the randomized procedure, the expected number of participants that show consistency is also close to two, although the specific participants which are consistent vary with the realization of the randomized procedure. More specifically, after 1000 runs of the randomized procedure, the mean number of consistent participants we observed was 2.33.

 \begin{table}[!tb]
\centering
 \begin{tabular}{||c| c|| c| c ||} 
 \hline
 participant & Confidence interval & participant & Confidence interval \\ [0.5ex] 
 \hline\hline
 1 & \cellcolor[RGB]{175,234,180}(0.03,\text{ }48.59) & 6 & \cellcolor[RGB]{175,234,180} (0.04,\text{ }3.75) \\ 
 2 & \cellcolor[RGB]{175,234,180} (0.03,\text{ }137.58) & 7  &\cellcolor[RGB]{175,234,180} (0.01,\text{ }0.50)\\
 3 & \cellcolor[RGB]{175,234,180} (0.01,\text{ }3.63) & 8 &\cellcolor[RGB]{255, 182, 193}(1.98,\text{ }518130.21)\\
 4 & \cellcolor[RGB]{175,234,180}(0.49,\text{ }63.48) & 9  & \cellcolor[RGB]{255, 182, 193}(0.76,\text{ }77.24)\\
 5 &\cellcolor[RGB]{175,234,180} (0.46,\text{ }1.37) & 10  &\cellcolor[RGB]{255, 182, 193}(0.26,\text{ }48758.33)\\
 \hline
 \end{tabular}
 \caption{Experimental results for Jarzynski's equality. We include the confidence intervals for the left hand side of \eqref{prediction II}, which we obtain after bootstrapping the observed values of $\Delta E_{ext}(\vect{x})$ for the forward process 1000 times and estimating $\langle e^{-\beta \Delta E_{ext}(\vect{X})}\rangle$ by its mean for each set of bootstrapped data.
 In our experiment we have $\Delta F=0$ in the right hand side of \eqref{prediction II}, resulting in a theoretical prediction of $\langle e^{-\beta \Delta E_{ext}(\vect{X})}\rangle=1.0$. Note, that for most subjects the value of $1.0$ lies inside the confidence interval, which does not hold when assuming unsuitable loss functions, as discussed at the end of the Results. Participants that achieve at least $50\%$ adaptation (c.f. Figure \ref{hyste plot}) are shaded by a green background color.}
 \label{jarz participants}
 \end{table}


\section{Discussion}


In our experiment we have investigated the hypothesis that human sensorimotor adaptation may be participant to the thermodynamic fluctuation theorems first reported by Crooks \cite{crooks1999entropy} and Jarzynski \cite{jarzynski2000hamiltonian}. In particular, we tested whether changes in sensorimotor error induced externally by an experimental protocol are linearly related to the log-ratio of the  probabilities of behavioral trajectories under a given forward and time-reversed backward protocol of a sequence of visuomotor rotations. We found that participants' data, in all cases where participants showed an appropriate adaptive response, was consistent with this prediction
or close to its confidence interval bounds, as expected from our simulations with finite sample size.
Moreover, we found that the exponentiated error averaged over the path probabilities was statistically compatible with unity for these participants, in line with Jarzynski's theorem. 

Together these results not only extend the experimental evidence of \linebreak Boltzmann-like relationships between the probabilities of behavior and the corresponding order-inducing functions---such as energy, utility, or sensorimotor error---from the equilibrium to the non-equilibrium domain, but also from simple physical systems to more complex learning systems when studying adaptation in changing environments, deepening, thus, the parallelism between thermodynamics in physics and decision-making systems \cite{ortega2013thermodynamics}.

When testing for the validity of thermodynamic relations, one of the most critical issues is the choice of the energy function, that is, in our case, the error cost function. In physical systems, the energy function is usually hypothesized following from simple models involving point masses, springs, rigid bodies, etc., and generally requires knowledge of the degrees of freedom of the system under consideration. Here we have used an exponential quadratic error as a utility function, as it has been suggested previously that human pointing behavior can be best captured by loss functions that approximately follow a negative parabola for small errors and then level off for large errors \cite{kording2004loss}. In the absence of very large errors, many studies in the literature on sensorimotor learning have only used the quadratic loss term  \cite{wolpert1995internal,todorov2002optimal}. Quadratic errors have also been advocated in the context of the central limit theorem and in terms of prediction errors in the context of predictive coding \cite{rao1999predictive,todorov2008general,still2009information,still2012thermodynamics}. Thus, our assumptions regarding the loss function are compatible with the literature at large. Crucially, the reported results fail when assuming non-sensical cost functions, like the Mexican hat.

Experimental tests of both Jarzynski's equality \eqref{prediction II} and Crooks fluctuation theorem \eqref{prediction} have been previously reported  in classical physics \cite{douarche2005experimental,collin2005verification,toyabe2010experimental,saira2012test,liphardt2002equilibrium} and also, in the case of Jarzynski's equality, in quantum physics \cite{an2015experimental,smith2018verification}. Importantly, these results have been successfully tested in several contexts: unfolding and refolding processes involving RNA \cite{collin2005verification,liphardt2002equilibrium}, electronic transitions between electrodes manipulating a charge parameter \cite{saira2012test}, rotation of a macroscopic object inside a fluid surrounded by magnets where the current of a wire attached to the macroscopic object is manipulated \cite{douarche2005experimental}, and a trapped ion \cite{an2015experimental,smith2018verification}. Despite differences in physical realization, protocols, and energy functions (and thus work functions), all the above experiments follow the same basic design behind the approach presented here. This supports the claim that 
fluctuation theorems do not necessarily rely on involved physical assumptions but are simple mathematical properties of certain stochastic processes \cite{hack2022jarzyskis}, although originally they were derived in the context of non-equilibrium thermodynamics \cite{jarzynski1997equilibrium,crooks1998nonequilibrium}. 

Mathematically, Crooks theorem  \eqref{prediction} holds for any Markov process (i), whose initial distribution is in equilibrium (ii), and whose transition probabilities satisfy detailed balance with respect to the corresponding equilibrium distributions (iii) \cite{hack2022jarzyskis}. 
Our experimental test of Equation~\eqref{prediction} can be seen, thus, as a test for the hypothesis that human sensorimotor adaptation processes satisfy conditions (i), (ii), and (iii). Condition (i) requires adaptation to be  Markovian, which is in line with most error-driven models of sensorimotor adaptation  \cite{shadmehr2012} that assume some internal state update of the form $x_{t+1}=f(x_t, e)$ with adaptive state $x$ and error $e$. While such models have proven fruitful for simple adaptation tasks like ours, they also have clear limitations, for example when it comes to meta-learning processes that have been reported in more complex learning scenarios \cite{braun2010,lieder2019}. Condition (ii) is supported by our data in the second and last rows of Figure \ref{plateaus all},
where it can be seen that participants' behavior at the beginning of each cycle is at least approximately consistent with the equilibrium behavior recorded prior to the start of the experiment. Condition (iii) requires that the adaptive process converges to the equilibrium distribution \eqref{Boltzmann} dictated by the environment and that the behaviour remains statistically unchanged when staying in that environment. Moreover, it requires that the equilibrium behavior at each energy level is time-reversible, that means, once adaption has ceased the trial-by-trial behavior would have the same statistics when played forward or backward in a video recording. Note, however, that does not imply time-reversibility over the entire adaptation trajectory, but is only required locally for each transition step. In our sensorimotor setting, this would mean that after a suitably long adaptation time with perfect adaptation there would ultimately be no hysteresis, and accordingly it would be impossible to tell where the learner has come from. If we regard, for example, Metropolis-Hastings as a plausible model of adaptation, as some kind of stochastic optimization scheme, detailed balance and time reversibility would be fulfilled \cite{SANBORN2016883,grau2018non}. What kind of model describes human adaptive behavior best, and whether such a model is compatible with detailed balance is ultimately an open question. In our experiment at least, the condition seems to be fulfilled well enough to stay within the confidence intervals associated with the predictions made by Crooks' theorem.

While Jarzynski's equality \eqref{prediction II} directly follows from Crooks theorem, weaker assumptions are sufficient to derive it \cite{hack2022jarzyskis,jarzynski1997equilibrium}. In particular, condition (iii) regarding detailed balance is not necessary, as it is only required that the behavioral distribution does not change anymore once the equilibrium distribution is reached. Thus, Equation~\eqref{prediction II} can be used as a test for the weaker hypothesis that human sensorimotor adaptation satisfies conditions (i), (ii) and stationarity after convergence. While Jarzynski's equality only requires samples from the forward process, Crooks theorem also tests the relation between the forward and the backward processes. In particular, Crooks theorem
decouples the information processing with respect to any particular environment from the biases introduced by the adaptation history, that is, it assumes the transition probabilities for any given environment are independent of the history. In other words, the conditional probabilities have no memory and, thus, all memory effects are explained in terms of the state of the learning system prior to making some decision. Hence, the observed difference in behaviour after having adapted to the same environment, the hysteresis, is solely explained in terms of the information processing history before encountering the environment. Such hysteresis effects are not only common in simple physical systems like magnets or elastic bands, but have also been reported for sensorimotor tasks \cite{kelso1994,schack2011,turnham2012facilitation}. The hysteresis effects we report in Figure~\ref{hyste plot} are in line with a system obeying Crooks theorem and can be replicated using Markov Chain Monte Carlo simulations of adaptation \cite{grau2018non}.

Our study is part of a number of recent studies that have tried to harness equilibrium and non-equilibrium thermodynamics to gain new theoretical insights into simple learning systems \cite{goldt2017stochastic,perunov2016statistical,england2015dissipative,still2012thermodynamics,ortega2013thermodynamics}.
For example, the information that can be acquired by learning in simple forward neural networks has been shown to be bounded by thermodynamic costs given by the entropy change in the weights and the heat dissipated into the environment \cite{seifert2012stochastic}. More generally, when interpreting a system's response to a stochastic driving signal in terms of computation, the amount of non-predictive information contained in the state about past environmental fluctuations is directly related to the amount of thermodynamic dissipation \cite{still2012thermodynamics}. This suggests that thermodynamic fundamentals, like the second law, can be carried over to learning systems.
Consider, for example, a Bayesian learner where the utility is given by the log-likelihood model and where the data are presented either in one chunk for a single update, or consecutively in little batches with many little updates. Rather than having one big surprise, in the latter case the cumulative surprise is much smaller as prior expectations can be continuously adapted, up to a point where the cumulative surprise reaches a lower bound given by the log-likelihood of the data, which corresponds to the free energy difference before and after learning \cite{grau2018non}. 
Fluctuation theorems have recently also been attributed a fundamental role in the context of the Free Energy Principle, with relations to information geometry and decision-theoretic concepts like risk, ambiguity, expected information gain and expected value
\cite{parr2020markov,da2021bayesian}. Due to the central role of the concept of variational free energy in inference processes \cite{GottwaldBraun2020}, this raises the interesting question in how far our results may generalise to any belief-updating process, including for example perceptual inference and perceptual hysteresis.
Finally, it has even been suggested that the dissipation of absorbed work as it is studied in a generalized Crooks theorem may underlie a general thermodynamic mechanism for self-organization and adaptation in living matter \cite{england2015dissipative}, raising the question of whether such a general principle of adaptive dissipation could also govern biological learning processes \cite{perunov2016statistical}.

\bibliographystyle{plain}
\bibliography{main.bib}

\begin{thebibliography}{10}

\bibitem{an2015experimental}
Shuoming An, Jing-Ning Zhang, Mark Um, Dingshun Lv, Yao Lu, Junhua Zhang,
  Zhang-Qi Yin, HT~Quan, and Kihwan Kim.
\newblock Experimental test of the quantum jarzynski equality with a
  trapped-ion system.
\newblock {\em Nature Physics}, 11(2):193--199, 2015.

\bibitem{braun2010}
Daniel~A. Braun, Carsten Mehring, and Daniel~M. Wolpert.
\newblock Structure learning in action.
\newblock {\em Behavioural Brain Research}, 206(2):157--165, 2010.

\bibitem{chib1995understanding}
Siddhartha Chib and Edward Greenberg.
\newblock Understanding the metropolis-hastings algorithm.
\newblock {\em The american statistician}, 49(4):327--335, 1995.

\bibitem{cohen2004note}
EGD Cohen and David Mauzerall.
\newblock A note on the jarzynski equality.
\newblock {\em Journal of Statistical Mechanics: Theory and Experiment},
  2004(07):P07006, 2004.

\bibitem{collin2005verification}
Delphine Collin, Felix Ritort, Christopher Jarzynski, Steven~B Smith, Ignacio
  Tinoco, and Carlos Bustamante.
\newblock Verification of the crooks fluctuation theorem and recovery of rna
  folding free energies.
\newblock {\em Nature}, 437(7056):231--234, 2005.

\bibitem{crooks1998nonequilibrium}
Gavin~E Crooks.
\newblock Nonequilibrium measurements of free energy differences for
  microscopically reversible markovian systems.
\newblock {\em Journal of Statistical Physics}, 90(5):1481--1487, 1998.

\bibitem{crooks1999entropy}
Gavin~E Crooks.
\newblock Entropy production fluctuation theorem and the nonequilibrium work
  relation for free energy differences.
\newblock {\em Physical Review E}, 60(3):2721, 1999.

\bibitem{crooks2000path}
Gavin~E Crooks.
\newblock Path-ensemble averages in systems driven far from equilibrium.
\newblock {\em Physical review E}, 61(3):2361, 2000.

\bibitem{da2021bayesian}
Lancelot Da~Costa, Karl Friston, Conor Heins, and Grigorios~A Pavliotis.
\newblock Bayesian mechanics for stationary processes.
\newblock {\em Proceedings of the Royal Society A}, 477(2256):20210518, 2021.

\bibitem{de2013non}
Sybren~Ruurds De~Groot and Peter Mazur.
\newblock {\em Non-equilibrium thermodynamics}.
\newblock Courier Corporation, 2013.

\bibitem{douarche2005experimental}
Fr{\'e}d{\'e}ric Douarche, Sergio Ciliberto, Artyom Petrosyan, and Ivan
  Rabbiosi.
\newblock An experimental test of the jarzynski equality in a mechanical
  experiment.
\newblock {\em EPL (Europhysics Letters)}, 70(5):593, 2005.

\bibitem{england2015dissipative}
Jeremy~L England.
\newblock Dissipative adaptation in driven self-assembly.
\newblock {\em Nature nanotechnology}, 10(11):919--923, 2015.

\bibitem{goldt2017stochastic}
Sebastian Goldt and Udo Seifert.
\newblock Stochastic thermodynamics of learning.
\newblock {\em Physical review letters}, 118(1):010601, 2017.

\bibitem{gottwald2019}
Sebastian Gottwald and Daniel~A. Braun.
\newblock Bounded rational decision-making from elementary computations that
  reduce uncertainty.
\newblock {\em Entropy}, 21(4), 2019.

\bibitem{GottwaldBraun2020}
Sebastian Gottwald and Daniel~A. Braun.
\newblock The two kinds of free energy and the bayesian revolution.
\newblock {\em PLOS Computational Biology}, 16(12):1--32, 12 2020.

\bibitem{grau2018non}
Jordi Grau-Moya, Matthias Kr{\"u}ger, and Daniel~A Braun.
\newblock Non-equilibrium relations for bounded rational decision-making in
  changing environments.
\newblock {\em Entropy}, 20(1):1, 2018.

\bibitem{lieder2019}
Thomas~L Griffiths, Frederick Callaway1, Michael~B Chang, Erin Grant, Paul~M
  Krueger, and Falk Lieder.
\newblock Doing more with less: meta-reasoning and meta-learning in humans and
  machines.
\newblock {\em Current Opinion in Behavioral Sciences}, 29:24--30, 2019.

\bibitem{hack2022jarzyskis}
Pedro Hack, Sebastian Gottwald, and Daniel~A. Braun.
\newblock Jarzyski's equality and crooks' fluctuation theorem for general
  markov chains.
\newblock {\em arXiv preprint arXiv:2202.05576}, 2022.

\bibitem{jarzynski1997equilibrium}
Christopher Jarzynski.
\newblock Equilibrium free-energy differences from nonequilibrium measurements:
  A master-equation approach.
\newblock {\em Physical Review E}, 56(5):5018, 1997.

\bibitem{jarzynski2000hamiltonian}
Christopher Jarzynski.
\newblock Hamiltonian derivation of a detailed fluctuation theorem.
\newblock {\em Journal of Statistical Physics}, 98(1):77--102, 2000.

\bibitem{jarzynski2004nonequilibrium}
Christopher Jarzynski.
\newblock Nonequilibrium work theorem for a system strongly coupled to a
  thermal environment.
\newblock {\em Journal of Statistical Mechanics: Theory and Experiment},
  2004(09):P09005, 2004.

\bibitem{jarzynski2011equalities}
Christopher Jarzynski.
\newblock Equalities and inequalities: Irreversibility and the second law of
  thermodynamics at the nanoscale.
\newblock {\em Annu. Rev. Condens. Matter Phys.}, 2(1):329--351, 2011.

\bibitem{kelso1994}
J.J. Kelso, J.A.S.and~Buchanan and T.~Murata.
\newblock Multifunctionality and switching in the coordination dynamics of
  reaching and grasping. human movement science.
\newblock {\em Current Opinion in Behavioral Sciences}, 13:63--94, 1994.

\bibitem{kording2004loss}
Konrad~Paul K{\"o}rding and Daniel~M Wolpert.
\newblock The loss function of sensorimotor learning.
\newblock {\em Proceedings of the National Academy of Sciences},
  101(26):9839--9842, 2004.

\bibitem{lieb1991}
Elliott~H. Lieb and Jakob Yngvason.
\newblock The physics and mathematics of the second law of thermodynamics.
\newblock {\em Physics Reports}, 310(1):1--96, 1999.

\bibitem{cecilia2019}
Cecilia Lindig-León, Sebastian Gottwald, and Daniel~A. Braun.
\newblock Analyzing abstraction and hierarchical decision-making in absolute
  identification by information-theoretic bounded rationality.
\newblock {\em Frontiers in Neuroscience}, 13, 2019.

\bibitem{cecilia2021}
Cecilia Lindig-León, Gerrit Schmid, and Braun~Daniel A.
\newblock Bounded rational response equilibria in human sensorimotor
  interactions.
\newblock {\em Proc. R. Soc. B.}, 288(1962):20212094, 2021.

\bibitem{liphardt2002equilibrium}
Jan Liphardt, Sophie Dumont, Steven~B Smith, Ignacio Tinoco, and Carlos
  Bustamante.
\newblock Equilibrium information from nonequilibrium measurements in an
  experimental test of jarzynski's equality.
\newblock {\em Science}, 296(5574):1832--1835, 2002.

\bibitem{loschmidt1876ueber}
Joseph Loschmidt.
\newblock Ueber den zustand des w{\"a}rmegleichgewichtes eines system von
  k{\"o}rpern.
\newblock {\em Akademie der Wissenschaften, Wien.
  Mathematisch-Naturwissenschaftliche Klasse, Sitzungsberichte}, 73:128--135,
  1876.

\bibitem{mascollel1995}
Andreu Mas-Colell, Michael Whinston, and Jerry Green.
\newblock {\em Microeconomic theory}.
\newblock Oxford University Press, 1995.

\bibitem{ortega2013thermodynamics}
Pedro~A Ortega and Daniel~A Braun.
\newblock Thermodynamics as a theory of decision-making with
  information-processing costs.
\newblock {\em Proceedings of the Royal Society A: Mathematical, Physical and
  Engineering Sciences}, 469(2153):20120683, 2013.

\bibitem{park2003free}
Sanghyun Park, Fatemeh Khalili-Araghi, Emad Tajkhorshid, and Klaus Schulten.
\newblock Free energy calculation from steered molecular dynamics simulations
  using jarzynski’s equality.
\newblock {\em The Journal of chemical physics}, 119(6):3559--3566, 2003.

\bibitem{parr2020markov}
Thomas Parr, Lancelot Da~Costa, and Karl Friston.
\newblock Markov blankets, information geometry and stochastic thermodynamics.
\newblock {\em Philosophical Transactions of the Royal Society A},
  378(2164):20190159, 2020.

\bibitem{parrondo2015thermodynamics}
Juan~MR Parrondo, Jordan~M Horowitz, and Takahiro Sagawa.
\newblock Thermodynamics of information.
\newblock {\em Nature physics}, 11(2):131--139, 2015.

\bibitem{perunov2016statistical}
Nikolay Perunov, Robert~A Marsland, and Jeremy~L England.
\newblock Statistical physics of adaptation.
\newblock {\em Physical Review X}, 6(2):021036, 2016.

\bibitem{rao1999predictive}
Rajesh~PN Rao and Dana~H Ballard.
\newblock Predictive coding in the visual cortex: a functional interpretation
  of some extra-classical receptive-field effects.
\newblock {\em Nature neuroscience}, 2(1):79--87, 1999.

\bibitem{saira2012test}
O-P Saira, Y~Yoon, T~Tanttu, Mikko M{\"o}tt{\"o}nen, DV~Averin, and Jukka~P
  Pekola.
\newblock Test of the jarzynski and crooks fluctuation relations in an
  electronic system.
\newblock {\em Physical review letters}, 109(18):180601, 2012.

\bibitem{SANBORN2016883}
Adam~N. Sanborn and Nick Chater.
\newblock Bayesian brains without probabilities.
\newblock {\em Trends in Cognitive Sciences}, 20(12):883--893, 2016.

\bibitem{Schach2018}
Sonja Schach, Sebastian Gottwald, and Daniel~A. Braun.
\newblock Quantifying motor task performance by bounded rational decision
  theory.
\newblock {\em Frontiers in Neuroscience}, 12, 2018.

\bibitem{schack2011}
Christoph Schütz, Matthias Weigelt, Dennis Odekerken, Timo Klein-Soetebier,
  and Thomas Schack.
\newblock Motor control strategies in a continuous task space.
\newblock {\em Motor Control}, 15(3):321 -- 341, 2011.

\bibitem{seifert2005entropy}
Udo Seifert.
\newblock Entropy production along a stochastic trajectory and an integral
  fluctuation theorem.
\newblock {\em Physical review letters}, 95(4):040602, 2005.

\bibitem{seifert2012stochastic}
Udo Seifert.
\newblock Stochastic thermodynamics, fluctuation theorems and molecular
  machines.
\newblock {\em Reports on progress in physics}, 75(12):126001, 2012.

\bibitem{shadmehr2012}
Reza Shadmehr and Sandro Mussa-Ivaldi.
\newblock {\em Biological learning and control : how the brain builds
  representations, predicts events, and makes decisions}.
\newblock MIT Press, 2012.

\bibitem{smith2018verification}
Andrew Smith, Yao Lu, Shuoming An, Xiang Zhang, Jing-Ning Zhang, Zongping Gong,
  HT~Quan, Christopher Jarzynski, and Kihwan Kim.
\newblock Verification of the quantum nonequilibrium work relation in the
  presence of decoherence.
\newblock {\em New Journal of Physics}, 20(1):013008, 2018.

\bibitem{still2009information}
Susanne Still.
\newblock Information-theoretic approach to interactive learning.
\newblock {\em EPL (Europhysics Letters)}, 85(2):28005, 2009.

\bibitem{still2012thermodynamics}
Susanne Still, David~A Sivak, Anthony~J Bell, and Gavin~E Crooks.
\newblock Thermodynamics of prediction.
\newblock {\em Physical review letters}, 109(12):120604, 2012.

\bibitem{todorov2008general}
Emanuel Todorov.
\newblock General duality between optimal control and estimation.
\newblock In {\em 2008 47th IEEE Conference on Decision and Control}, pages
  4286--4292. IEEE, 2008.

\bibitem{todorov2002optimal}
Emanuel Todorov and Michael~I Jordan.
\newblock Optimal feedback control as a theory of motor coordination.
\newblock {\em Nature neuroscience}, 5(11):1226--1235, 2002.

\bibitem{toyabe2010experimental}
Shoichi Toyabe, Takahiro Sagawa, Masahito Ueda, Eiro Muneyuki, and Masaki Sano.
\newblock Experimental demonstration of information-to-energy conversion and
  validation of the generalized jarzynski equality.
\newblock {\em Nature physics}, 6(12):988--992, 2010.

\bibitem{turnham2012facilitation}
Edward~JA Turnham, Daniel~A Braun, and Daniel~M Wolpert.
\newblock Facilitation of learning induced by both random and gradual
  visuomotor task variation.
\newblock {\em Journal of Neurophysiology}, 107(4):1111--1122, 2012.

\bibitem{wkeglarczyk2018kernel}
Stanislaw Weglarczyk.
\newblock Kernel density estimation and its application.
\newblock In {\em ITM Web of Conferences}, volume~23. EDP Sciences, 2018.

\bibitem{wolpert1995internal}
Daniel~M Wolpert, Zoubin Ghahramani, and Michael~I Jordan.
\newblock An internal model for sensorimotor integration.
\newblock {\em Science}, 269(5232):1880--1882, 1995.

\bibitem{ytreberg2004efficient}
F~Marty Ytreberg and Daniel~M Zuckerman.
\newblock Efficient use of nonequilibrium measurement to estimate free energy
  differences for molecular systems.
\newblock {\em Journal of computational chemistry}, 25(14):1749--1759, 2004.

\end{thebibliography}

\appendix
\section{Appendix: Methods}

\subsection{Theoretical methods}
\label{theo meth}

The derivation of \eqref{prediction} and \eqref{prediction II} in the context of general Markov chains can be found in \cite{hack2022jarzyskis}. A similar proof of \eqref{prediction II} under stronger assumptions was derived in \cite{crooks1998nonequilibrium} and a different one using the same assumptions was given in \cite{jarzynski1997equilibrium}. Regarding \eqref{prediction}, a similar proof can be found in \cite{crooks1998nonequilibrium}. Note, however, that the usual definition of work in thermodynamics is slightly different for the forward and backward process, based on the physical definition of time reversal and the associated symmetry for the work values. In our case, we define the driving signal that is analogous to the work concept in the same way, for both forward and backward process. In this case, for Equation~\eqref{prediction} to hold, we need to assume that $E_1=E_0$  both in the forward and backward process \cite{hack2022jarzyskis}. Fortunately, this is true for our protocol, since we begin both forward and backward protocol with some washout trials without perturbation.
It should also be pointed out that, in order for the elements involved in Jarzynski's and Crooks' derivations to be well-defined, the equilibrium probability density associated to each step in the Markov chain ought to be non-zero at both the starting and ending point of that step \cite{hack2022jarzyskis}. This will play a relevant role in the choice of the support $A_n$ for the equilibrium distributions $p_n^{eq}$ in Section \ref{exp design}.

\begin{figure}[!tb]
\centering
    \includegraphics[scale=0.21]{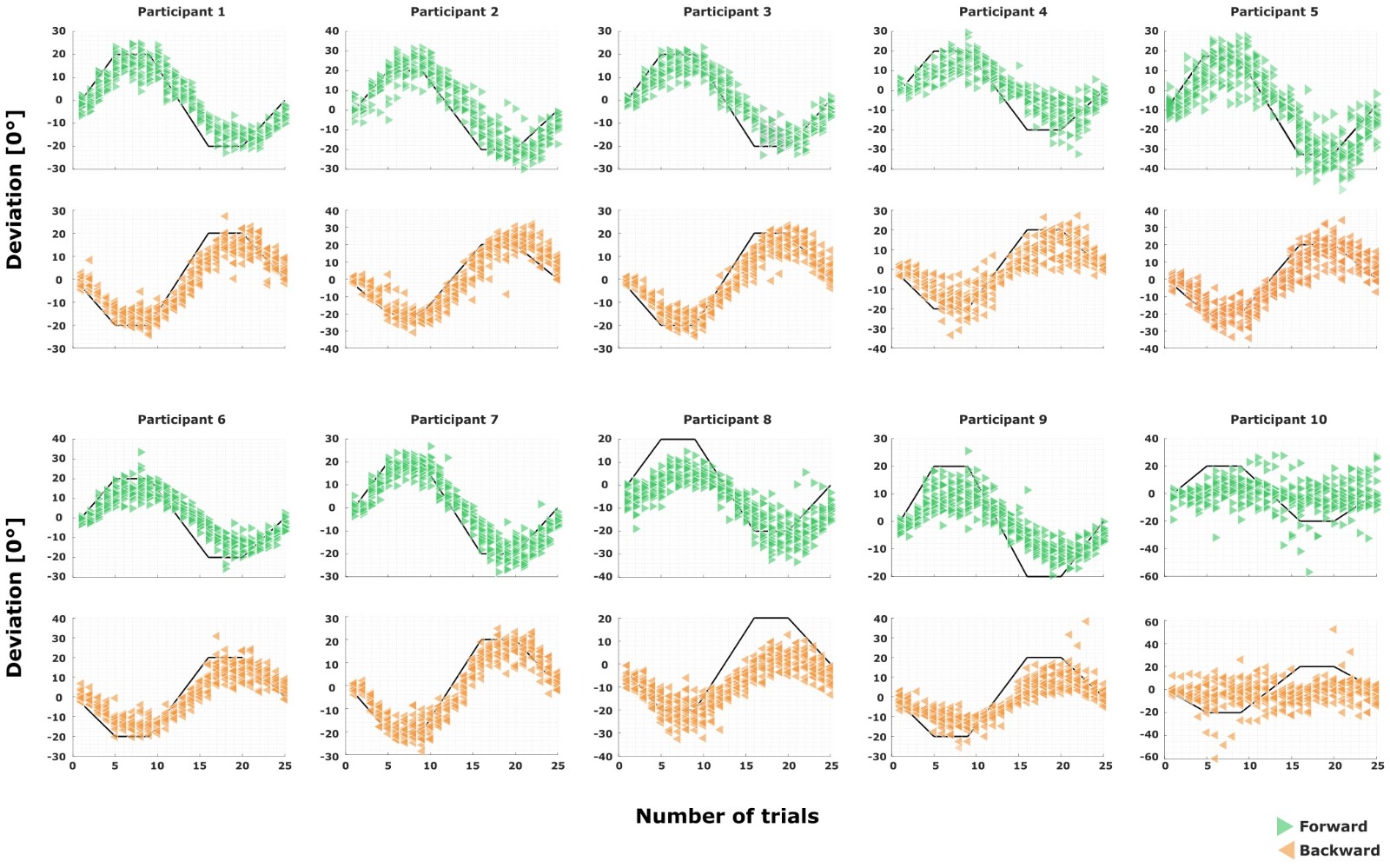}
    \caption{Observed angles in the forward and backward processes. The black line represents the protocol while the filled triangles correspond to both the forward trajectories, first and third rows in green, and the backward trajectories, second and fourth rows in red.
    }
    \label{forward all}
\end{figure}

\subsection{Simulation methods}
\label{simu meth}
In this section, we explain in detail how we simulated \eqref{prediction} and \eqref{prediction II}.

\subsubsection{Metropolis-Hastings algorithm}
\label{metro-has}
We use \cite{chib1995understanding} as reference for this section. However, for simplicity, we skip over several technical details and may oversimplify some notions.

The Metropolis-Hastings algorithm is a procedure which allows to obtain samples $x$ from a distribution $p$ that is proportional to some function $f$, that is, $p(x) = \frac{1}{Z} f(x)$. There are three concepts relevant to this algorithm: $U$, $q$ and $\alpha$. They are defined as follows
	\begin{itemize}
	\item $U(A)$ stands for the uniform distribution over some set $A \subseteq \mathbb R$. 
	\item $q(\cdot,\cdot)$ is called the \emph{candidate generating density}. The role of $q$ in the algorithm is to generate a new point $y$ given a previous point $x$, with $y$ being sampled from the distribution $q(x,\cdot)$. In our case we define the density function in $y$ with $\int_{-90}^{90} q(x,y) dy = 1$, as we assume that movements will be towards the target ($0^\circ$ direction) under a maximally induced error of $20^\circ$. Accordingly, we can expect that practically all responses will be covered by choosing a support of $\pm 90^\circ$.
	\item  $\alpha(\cdot,\cdot)$ is defined as follows:
	\begin{alignat*}{2}
		\alpha(x,y) &= \text{min} \Big\{ \frac{f(y)q(y,x)}{f(x)q(x,y)},1\Big\} &&\text{ if } f(x)q(x,y)>0,\\
		&= 1 &&\text{ otherwise} 
	\end{alignat*}
	and is included in the algorithm as a filter on the samples proposed by $q$, so that some of these samples will be accepted and some will be rejected, to make the samples appear to be sampled from $p$.
\end{itemize}

We can now introduce the Metropolis-Hastings algorithm. The algorithm is initialized at an arbitrary value $x_0$ and then repeats the following steps for $i = 1,2,..,M$:  
\begin{enumerate}[label=(\roman*)]
    \item Generate $y$ from $q(x_{i-1},\cdot)$ and $u$ from $U(0, 1)$.
    \item If $u \leq \alpha(x_{i-1},y)$, then $x_i=y$.
    \item Otherwise, $x_i=x_{i-1}$.
\end{enumerate}
Finally, the algorithm returns the values $(x_1,..,x_M)$.

Note that the density of transitions from $x$ to $y$ is therefore given by
\begin{equation*}
    p_M(x,y)=q(x,y) \alpha(x,y) \quad \text{ if } x \neq y ,
\end{equation*}
which satisfies detailed balance with respect to $p\propto f$ \cite{chib1995understanding}. Thus, $p$ is the stationary distribution of the resulting Markov process, and so the $x_i$ can be regarded samples from $p$ after the chain has passed a transient stage after which the effect of the initialization is negligible. Notice, in our implementation, described below, we only require the burn-in phase for the initial energy in order to make sure that the process starts in the corresponding stationary state. However, since we are interested in the adaption process during a changing energy signal, we only use the first sample ($M=1$) for the remaining steps, conditioned on the sample from the previous step.

\begin{figure}[!tb]
\centering
   \includegraphics[scale=0.2]{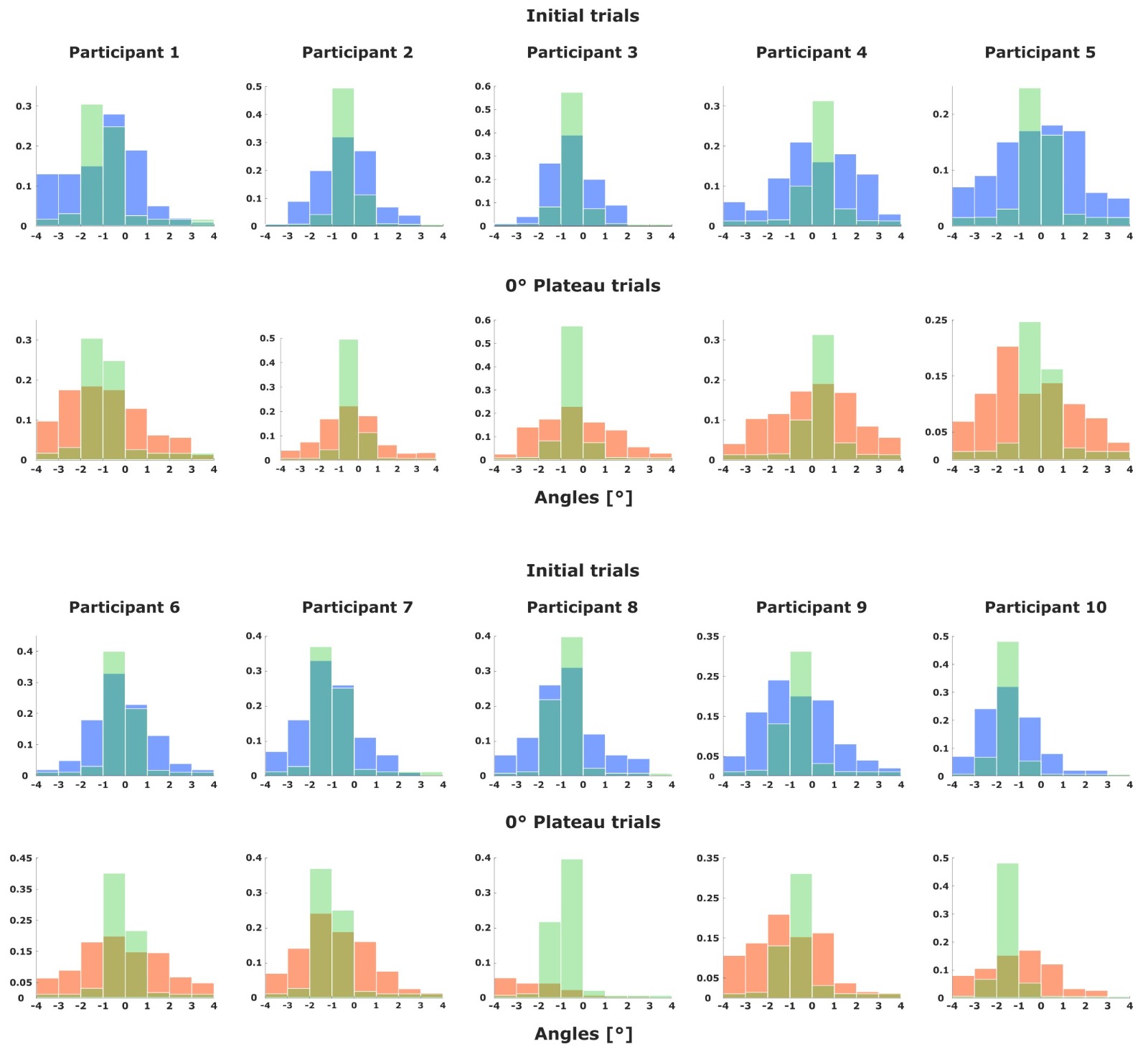}
   \caption{Comparison between participants' behaviour in washout trials (between perturbation cycles) with the fitted equilibrium distribution (recorded before participants experienced any perturbation). The first and third rows compare the normalized histogram of the angles observed during the initial 100 trials (blue color), with the histogram of the fitted equilibrium distribution~\eqref{Boltzmann} over the same trials (green color). The second and the forth rows compare the same fitted equilibrium distribution (green color) with the normalized histogram of the angles observed in the $0^\circ$ deviation plateaus (washout trials) which separate forward and backward protocol (red color). Note the plateau in each cycle consists of 10 points, from which we only include the last 8 to avoid large aftereffects. The application of Crooks' theorem requires that subjects fully equilibrate between protocols, that is, in our case their behavior in washout trials should return to the fitted equilibrium behavior at the start of the experiment. Compare the discussion on condition $(ii)$ on page 11.}
   \label{plateaus all}
\end{figure}

\subsubsection{Implementation}
Given a set of equilibrium distributions $(p_0,..,p_N)$, we use the Metropolis-Hastings algorithm on their proportional functions $(f_0,..,f_N)$ to generate two paths: the forward path where we apply the algorithm once at step $i$ ($M=1$ in Section \ref{metro-has}, as explained above), with $p=p_i$, and the backward path where we do the same but with the distributions in the reverse order. In particular, we consider 
\begin{equation}
\label{simu distri}
f_n(x) = e^{-E_n(x)}  
\end{equation}
for $n=0,..,24$ for the forward process, where, for $n=1,..,24$, we take
\begin{equation}
\label{ener simu}
    E_n(x)=-e^{-(x-\theta_n)^2}
\end{equation}
with $\theta_n$ given by \eqref{protocol}, and for $n=0$ we consider
\begin{equation}
\label{modified energy}
    E_0(x)=
    \begin{cases}
      -(x+2) &\text{if } x <-2, \\
      -e^{-x^2} &\text{if } -2\leq x \leq 2, \\
      x-2 &\text{if } 2<x.
    \end{cases}
\end{equation}
We will refer to the application of the algorithm following the sequence in \eqref{simu distri} with $M=1$ for each $n=1,..,25$ as a \emph{cycle}. Note $E_0$ in \eqref{modified energy} differs from $E_n$ in \eqref{ener simu} for $n=1,..,24$. While we would like to take $E_0$ as in  \eqref{ener simu} with $\theta_0=0$, since one of our hypothesis is the simulations sample the first point in each cycle from 
\begin{equation*}
p_0(x) \propto e^{e^{-x^2}},
\end{equation*}
the values of $p_0$ for $x \not \in [-2,2]$ are quite indistinguishable once we fix a certain precision. As a result, the algorithm does not converge to $p_0$ in the long run. To avoid this difficulty, we simply modify the function outside $[-2,2]$ such that points there become distinguishable. This results in the algorithm converging to a distribution close to $p_0$. Note this modification only applies to the generation of the initial samples, hence, we use \eqref{ener simu} to calculate $\Delta E_{ext}(\vect{x})$.

The candidate generating density we use for the $n$th step with $n=1,..,24$ is a normal distribution with mean equal to the $(n-1)$th sample and standard deviation equal to the mean of the distances between subsequent points in the observed data, which turns out to be around $5$. Using the values generated by the algorithm during a cycle, we calculate $\Delta E_{ext}(\vect{x})$ for the forward process via the utilities in \eqref{ener simu}, and, after generating several of them, we apply kernel density estimation (see Section \ref{kernel density}) to estimate $\rho^F$ in \eqref{prediction}. We proceed analogously to estimate $\rho^B$ and, finally, use the obtained values of $\Delta E_{ext}(\vect{x})$ for the forward process together with the estimates of $\rho^F$ and $\rho^B$ to test \eqref{prediction}. This test is done differently for the simulation with the large number of sample and that with a small number of them. For the larger one, we simply use the least squares method as the estimate of \eqref{prediction} (cf. Figure \ref{2 simus} \textbf{A}). For the smaller one, however, we produce 1000 bootstraps from the produced values of $\Delta E_{ext}(\vect{x})$ and find a confidence interval for \eqref{prediction} from the curves we obtain from the pair ($\rho^F$, $\rho^B$) for each bootstrap (cf. Figure \ref{2 simus} \textbf{B}).


\subsection{Experimental methods}
In this section, we explain the specifics of how we tested experimentally both \eqref{prediction} and \eqref{prediction II}.

\subsubsection{Participants}
Ten participants $P_1,..,P_{10}$, five females and five males, participated in this study. Three of the authors were among the participants ($P_1$, $P_2$ and $P_3$). All other
participants provided written informed consent for participation
and were remunerated with 10 Euros per hour. The participants were undergraduate and graduate students. The procedures were approved by the Ethics committee of Ulm University. All methods were performed in accordance with the relevant
guidelines and regulations.

\subsubsection{Setup}
The experiment was run on a vBOT. Each participant performed the task using the handle of the right arm of the vBOT, which was manipulated with the dominant hand. The participants had no direct view of the handle but of a screen where its position, altered according to a protocol we describe in the following, was represented by a cursor.

\subsubsection{Experimental design}
\label{exp design}
Participants were asked to reach the center of a yellow rounded target on the screen with the center of their cursor. To begin each trial, the participants were asked to place the cursor inside a rounded initial position whose center was 15 cm away from the target's center along the same vertical. Once the cursor crossed the horizontal containing the center of the target,
the target became green if participants successfully situated the center of the cursor inside the target and red otherwise. Once the target changed its color, participants were asked to return the cursor to the initial position to begin the following trial. While both the target and the initial position were at the same place each trial, the cursor did not represent the movement of the handle veridically each trial. In particular, after 100 trials where the cursor position and the handle coincided, there were 1420 trials divided in 20 cycles of 66 trials where the cursor position was determined by rotating the vector going from the center of the initial position to the handle's position.
The rotation angle $\theta_n$ for each $n$ in any cycle $n=0,..,65$ was
\begin{equation}
\label{protocol}
    \begin{cases}
      \theta_n = \alpha(n) &\text{if } n=0,..,24\\
    \theta_n=0 &\text{if } n=25,..,32\\
    \theta_n = \alpha(57-n) &\text{if } n=33,..,57\\
    \theta_n=0 &\text{if } n=58,..,65\\
     \end{cases}
\end{equation}
where all angles are in degrees and
\begin{equation*}
\begin{split}
\alpha=&(0,5,10,15,20,20,20,20,20,15,10,5,0,-5,-10,\\
&-15,-20,-20,-20,-20,-20,-15,-10,-5,0).
\end{split}
\end{equation*}

\begin{figure}[!tb]
\centering
\includegraphics[scale=0.2]{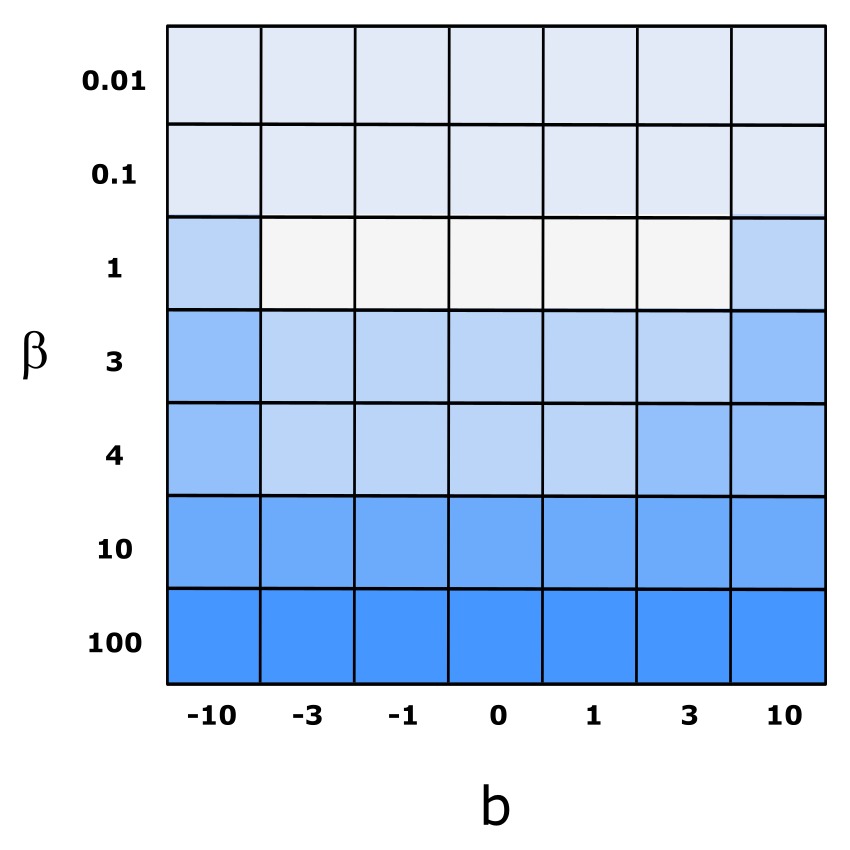}
\caption{Graphical representation of the accuracy of Crooks' fluctuation theorem for several pairs of parameters $(b,\beta)$, which we measure through $d_{b,\beta}$ as explained in Section \ref{model params}. The color intensity grows monotonically with the distance $d_{b,\beta}$ and is divided into six regions, namely, $d_{b,\beta} \leq 1$, $1<d_{b,\beta}\leq 3$, $3<d_{b,\beta} \leq 6$, $6< d_{b,\beta} \leq 11$, $11< d_{b,\beta} \leq 23$ and $23< d_{b,\beta}$. The actual values of $d_{b,\beta}$ can be found in Table \ref{distances}.}
\label{graph distances}
\end{figure}

For each $n=0,..,65$, we extract $\theta_n'$, the angle between the vertical segment joining the center of the initial position and the center of the target and the segment joining the center of the initial position and the handle in the first recorded point which is more than 12 cm apart from the center of the initial position. One can find the recorded angles $(x_0,..,x_{65})$ for both the forward and backward processes in Figure \ref{forward all}. For participant $P_j$, with $1 \leq j \leq 10$, we take $p_{n,j}=p^{eq}_n$ as the equilibrium distribution for the $n$-th trial,
where $b_j$ represents the bias introduced by the machine for participant $P_j$. We determine the bias as the mean of the initial 100 trials (where the cursor veridically represents the handle). The value $c_j$ represents the maximum deviation for participant $P_j$ among the distances $|x_n-(\theta_n+b_j)|$ and  $|x_{n-1}-(\theta_n+b_j)|$, 
which we use to fix the support of the equilibrium distribution for $P_j$ as 
$A_n = [\theta_n+b_j-c_j,\theta_n+b_j+c_j]$. The parameter $\beta_j$ represents the spread around the bias, which we pick once the bias and the support of the equilibrium distributions are fixed by requiring these distributions to maximize the likelihood of the observed values for the first 100 trials.
We observe the best spread parameters are between $\beta_j=0.25$ and $\beta_j=5$ for all participants. In order to choose the most suitable one for each participant, we consider the values between $0.25$ and $5$ that result from sequentially adding $0.25$ to the lowest value and pick as $\beta_j$ the one that maximizes the likelihood on the observed angles in the 100 initial trials--- see Figure \ref{plateaus all} for a comparison between the observed angles and the equilibrium distribution.
We discuss, in Section \ref{model params}, how the choice of the parameters $b_j$ and $\beta_j$ affect the results. Note the choice of $c_j$ does not directly affect how we measure the accuracy of the predictions, but is key in the maximum likelihood estimation of $\beta_j$.

Using the angles recorded during a cycle, we calculate $\Delta E_{ext}(\vect{x})$ via $p_{n,j}$ for both the forward and backward processes, and, using the 20 values per participant, we estimate $\rho^F$ and $\rho^B$ in \eqref{prediction} through kernel density estimation (see Section \ref{kernel density}). Finally, we bootstrap the obtained values of $\Delta E_{ext}(\vect{x})$ for the forward and backward process to obtain several estimates of $\rho^F$ and $\rho^B$. Each of these pairs is used to produce a curve that estimates \eqref{prediction}. The mean of these curves for each participant is what we compare to \eqref{prediction} in Figure \ref{together}. The same values of $\rho^F$ are used to test \eqref{prediction II} (cf. Table \ref{jarz participants}).

   \begin{table}[!tb]
\centering
 \begin{tabular}{||P{1cm}| P{1cm}| P{1cm}| P{1cm}| P{1cm} | P{1cm}|P{1cm} |P{1cm} ||} 
 \hline
 $b \text{ }\backslash \text{ }\beta$ & 0.01 & 0.1 & 1& 3  &4 &10 &100\\ [0.5ex] 
 \hline\hline
 -10 & 2.54 & 2.73 & 4.51 & 8.15& 10.15 & 22.53 & 202.52 \\
 -3 & 2.19 & 1.99 & 0.33 & 3.44& 5.42 & 17.79 & 197.80\\
 -1 & 2.42 & 2.26 & 0.48 & 3.08 & 4.81 & 17.57 &  197.55\\
 0 & 2.35 & 2.18 & 0.53 &3.31 & 5.02 & 17.63 & 197.61\\ 
 1 & 2.08 & 1.91  & 0.47 & 3.61 & 5.58 & 17.89 & 197.90\\
 3 & 1.66 & 1.48 & 0.51 & 4.22 & 6.22 & 18.31 & 198.33\\
 10 & 1.62 & 1.79 & 3.60 & 8.98 & 10.98 & 21.60 & 201.60 \\
 \hline
 \end{tabular}
 \caption{Mean distance between the theoretical prediction in \eqref{prediction} and the mean curve we obtain from bootstrapping the observed angles (see Section \ref{exp design}) for several pairs of parameters $(b,\beta)$. In particular, we consider the combinations having  $b=-10,-3,-1,0,1,3,10$ and $\beta=0.01,0.1,1,3,4,10,100$.}
 \label{distances}
 \end{table}

\subsubsection{Kernel density estimation}
\label{kernel density}
In order to determine the probability distributions $\rho^F$ and $\rho^B$ in \eqref{prediction}, we use \emph{kernel density estimation} \cite{wkeglarczyk2018kernel}. Kernel density estimation consists of choosing a function $K$, the \emph{kernel}, and a positive number $h>0$, the \emph{bandwidth}, and approximating $p$ by distributions of the form
\begin{equation*}
    \frac{1}{nh} \sum_{i=1}^n K\left(\frac{x-x_i}{h}\right).
\end{equation*}
We consider here $K$ to be a standard normal distribution. Notice we simply estimate $p$ as a sum of standard normal distributions around each observed point $x_i$, for $i=1,..,n$, and decide how much each $x_i$ influences other points in $\mathbb{R}$ via $h$. We fix $h=0.7$ throughout this work.

   \begin{table}[!tb]
\centering
 \begin{tabular}{||c| c|| c| c ||} 
 \hline
 $\lambda$ & Mean distance & $\lambda$&  Mean distance\\ [0.5ex] 
 \hline\hline
 0 & 11.68 & 0.75 & 6.04 \\ 
 0.25 & 10.78 & 1  & 5.02\\
 0.5 & 8.75 &  & \\
 \hline
 \end{tabular}
 \caption{Mean distance between the theoretical prediction in \eqref{prediction} and the mean curve we obtain from bootstrapping the observed angles (see Section \ref{exp design}) for several sensorimotor errors that are obtained as convex combinations of the exponential quadratic error \eqref{utility} and the Mexican hat \eqref{mex hat}. In particular, we consider sensorimotor errors of the form $\lambda f + (1-\lambda)g$, where $\lambda=0,0.25,0.5,0.75,1$, $f$ is the exponential quadratic error with $b=0$ and $\beta=4$ (which are close to the values fitted for the participants) and $g$ is the Mexican hat with $\sigma=4$. As expected, the mean distance diminishes as the weight of the exponential quadratic error increases.}
 \label{convex dist}
 \end{table}

\subsubsection{Robustness analysis}
\label{model params}

In this section, we measure model robustness using two approaches: $(i)$ using the exponential quadractic error \eqref{utility} and varying the parameters we fitted, i.e. $b$ and $\beta$, and $(ii)$ fixing a pair of parameters that are close to the optimal ones for each participant and taking convex combinations of the exponential quadractic error and the Mexican hat as sensorimotor errors.

\begin{figure}[tb!]
\centering
    \includegraphics[scale=0.2]{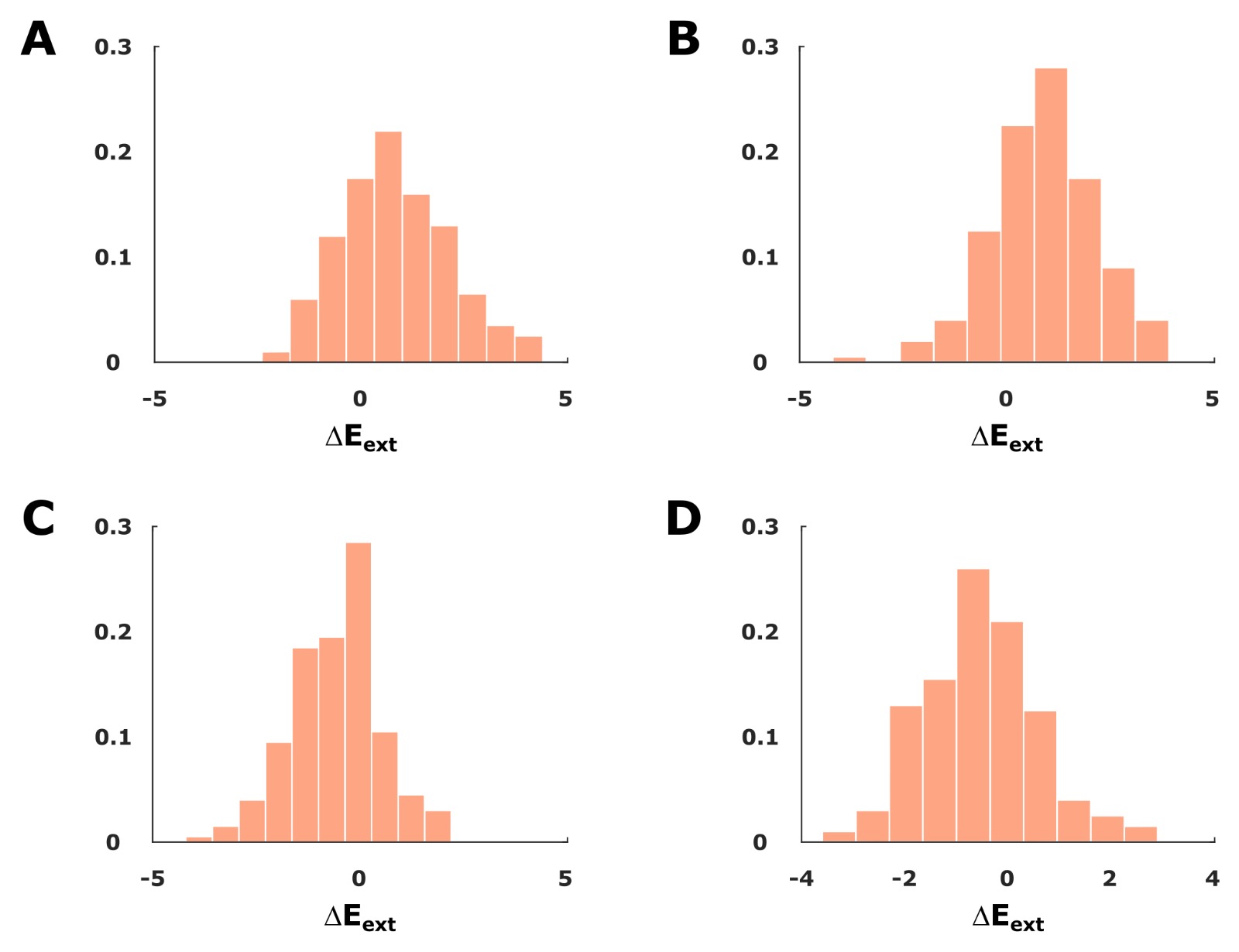}
    \caption{Histogram of the forward driving signal values using different values of $b$. In particular, we present the histograms for $b=1,-1,10,-10$. We include the first two since they are close to the values of $b$ we fit from the initial 100 trials (where no deviation is applied) and the last two to illustrate the grounds on which we discard certain parameter pairs. As expected from the observed hysteresis effect (cf. Figure \ref{hyste plot}), the histograms in \textbf{A} and \textbf{B}, which correspond to $b=1$ and $b=-1$, respectively, are biased towards positive values of the driving signal. When assuming implausible parameters, like the ones in \textbf{C} and \textbf{D}, which correspond to $b=10$ and $b=-10$, respectively, the bias shifts towards negative values (cf. \textbf{C} and \textbf{D}) and, even, shows a significant concentration of values around $0$ (cf. \textbf{C}). Note we observe, respectively, the same biases in the backward driving signals.}
    \label{work hist}
\end{figure}

As pointed out in Section \ref{exp design}, we fix the parameters in \eqref{utility} and \eqref{prediction}, via the initial 100 trials (where no perturbation is applied). To assess model robustness, we consider the effect of assuming the same model with different parameters. We consider, in particular, all pairs $(b,\beta)$ with $b \in \{-10,-3,-1,0,1,3,10\}$ and $\beta \in \{0.01,0.1,1,3,4,10,100\}$, since they cover a wide scope of the possible behaviour of \eqref{prediction} using the model in \eqref{utility}. For the robustness analysis we fit the data of all participants with the same parameter sets. In Figure \ref{work hist}, we show the histogram of the driving signals $\Delta E_{ext}(\mathbf{x})$ for different pairs of parameters $(b,\beta)$. Then, we follow the bootstrapping procedure from Section \ref{exp design} using the different values of $b$ and calculate the mean distance between the mean of the curves we obtain from the bootstraps and the theoretical prediction \eqref{prediction} with the different values of $\beta$. 
In particular, we consider the mean horizontal distance between the prediction and the mean curve at the points between $\Delta E_{ext}(\vect{x})=-4$ and $\Delta E_{ext}(\vect{x})=4$ (that is, the range of values of $\Delta E_{ext}(\vect{x})$ we present in Figure \ref{together}) with steps of $0.1$.
We denote the obtained mean distance as $d_{b,\beta}$.

To assess how well the parameters fit the data, we have to consider the plausibility of the data being generated by our model using the different parameter settings $(b,\beta)$.
Accordingly, it is not enough to simply look at $d_{b,\beta}$ as a goodness-of-fit measure. This is the case, as the underlying assumption in our model is that the data comes from a Markov chain where the equilibrium distributions at each step are given by the Boltzmann distribution \eqref{Boltzmann} with parameters $(b,\beta)$. In this situation, we expect participants to lag behind the utility they are adapting to most of the time, and hence, by definition, we expect the driving signal to be biased towards positive values. We can discard any parameter settings where this is not the case. Accordingly, we can disregard all pairs that have $b=10,-10$
---see Figure \ref{work hist}.
The values of $d_{b,\beta}$ for all pairs $(b,\beta)$ we considered can be found in Table \ref{distances} (see Figure \ref{graph distances} for a graphical comparison). As we can see there, the best parameters have $\beta=1$, $-3 \leq b \leq 3$, and mean distances which are both close to each other and significantly better than the rest. This was expected, since the hypothesis that the data observed at the plateaus follows \eqref{Boltzmann} for these parameters is not completely implausible (cf. Figure \ref{plateaus all}). The values $b \in \{-1,0,1\}$ and $\beta \in \{3,4\}$, which are the closest to the fitted parameters, also have a small mean distance (although larger than the best cases). In contrast, $d_{b,\beta}$ becomes significantly larger for the parameters that are clearly unlikely, that is, those that present a huge concentration of the probability around some point, i.e. the ones where the value of $\beta$ is large. In contrast, whenever the values of $\beta$ become small, the equilibrium distributions become all closer to a uniform and, although the mean distance does worsen when compared to the best cases, its values do not increase much. 

\begin{figure}[tb!]
\centering
    \includegraphics[scale=0.5]{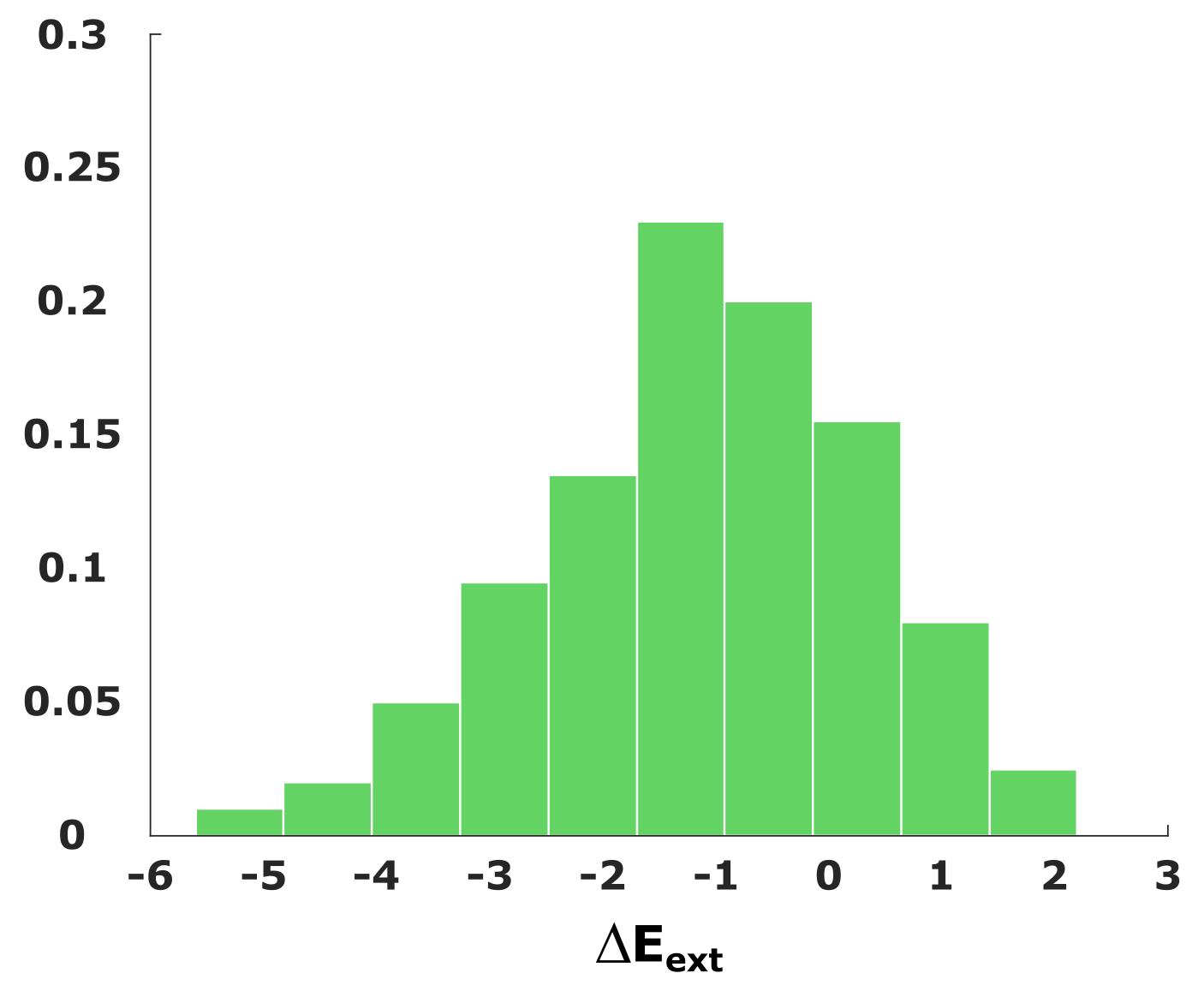}
    \caption{Histogram of the forward driving signals using an inverted Mexican hat \eqref{mex hat} with $\sigma=4$ as sensorimotor loss. Because of unexpected bias towards negative values of the driving signal we observe, it is unlikely the data was generated by a a Markov chain following such a sensorimotor loss and we can discard this model. Note we observe the same bias in the backward driving signals.}
    \label{mex histo}
    \end{figure}


To assess robustness with an obviously non-fitting utility, we consider an inverted Mexican hat as utility function, that is, we substitute \eqref{utility} by 

\begin{equation}
\label{mex hat}
   E'_n (x)= \frac{2}{\sqrt{3 \sigma} \pi^{\frac{1}{4}}} \bigg(1-\Big(\frac{x-\theta_n}{\sigma}\Big)^2\bigg) \exp \bigg( \frac{(x-\theta_n)^2}{2 \sigma^2} \bigg),
\end{equation}
where we take $\sigma=4$. In this scenario, the bootstrapped data does not reflect the trend of the theoretical prediction (cf. Figure \ref{togetherB}). Moreover, as illustrated in Figure \ref{mex histo}, the model presents an unexpected bias towards negative values of the driving signal.
Hence, as discussed above, the likelihood of the data coming from such a Markov chain is small and we can disregard this model. Furthermore, when following the same robustness analysis we performed on the pairs $(b,\beta)$ using the convex combinations $\lambda f + (1-\lambda) g$ as sensorimotor loss, where $\lambda=0,0.25,0.5,0.75,1$, $f$ is the exponential quadratic error with $b=0$ and $\beta=4$ (which are close to the values fitted for the participants) and $g$ is the Mexican hat with $\sigma=4$, we obtain that the mean distance decreases as $\lambda$ increases, as one can see in Table \ref{convex dist}.

\end{document}